\documentclass[11pt]{article}
\newif\ifzh
\zhfalse
\usepackage{algorithm,algpseudocode}
\usepackage{wrapfig}
\usepackage{needspace}
\usepackage{fvextra}
\PassOptionsToPackage{dvipsnames,table}{xcolor}
\usepackage{vendor/microsoft-tech-report}
\usepackage[T1]{fontenc}
\usepackage{graphicx,booktabs,array,tabularx,amsmath,amssymb}
\usepackage{flafter}
\usepackage[dvipsnames,table]{xcolor}
\usepackage{etoolbox}
\usepackage{url}
\usepackage{xurl}
\usepackage[unicode,colorlinks=true,linkcolor=black,citecolor=black,urlcolor=black]{hyperref}
\ifzh
  \usepackage{xeCJK}
  \setCJKsansfont{Heiti SC}
  \setCJKmonofont{Heiti SC}

  \renewenvironment{abstract}{\vskip.075in\centerline{\large 摘要}\vspace{0.5ex}\begin{quote}}{\par\end{quote}\vskip 1ex}
\fi
\definecolor{HCUBlue}{HTML}{0067D9}
\definecolor{WorseRed}{HTML}{B8322A}
\ifzh
  \definecolor{AXColor}{HTML}{176B68}
  \definecolor{GUIColor}{HTML}{986E18}
\else
  \definecolor{AXColor}{HTML}{004488}
  \definecolor{GUIColor}{HTML}{DDAA33}
\fi
\newcommand{\tlang}[2]{\ifzh#1\else#2\fi}

\newcommand{\capabilitytable}{%
\begin{table}[!htbp]
\caption{\tlang
{核心通道对照中的直接工具入口。勾选表示提供该类入口，不保证完整覆盖；其他入口仍可能间接重建相同能力。CUA 与原版系统的实际 schema 和权限另行记录，不冒用 HCU 通道标签。}
{Direct tool access in the channel and skill ablations. A check mark means that the configuration exposes that component directly.}}
\label{tab:capabilities}
\centering\small
\begin{tabular}{lcccc}
\toprule
\tlang{配置}{Configuration} & \tlang{直接代码/CLI}{Direct code/CLI} & AX & GUI & Skill\\
\midrule
HCU & $\checkmark$ & $\checkmark$ & $\checkmark$ & $\checkmark$\\
HCU w/o GUI & $\checkmark$ & $\checkmark$ & -- & $\checkmark$\\
HCU w/o AX & $\checkmark$ & -- & $\checkmark$ & $\checkmark$\\
HCU w/o skill & $\checkmark$ & $\checkmark$ & $\checkmark$ & --\\
\bottomrule
\end{tabular}
\end{table}
}

\newif\ifhcustatements
\hcustatementsfalse
\newcommand{\hcuappendixtitle}{\pdfbookmark[0]{Appendix}{appendix-part}%
  {\noindent\msftsans\bfseries\fontsize{20}{24}\selectfont\color{msftdark}Appendix\par}\vspace{0.4em}%
  {\noindent\color{msftline}\rule{\linewidth}{0.8pt}\par}\vspace{0.8em}}
\hypersetup{bookmarksnumbered=true}
\definecolor{BoxInk}{HTML}{1F2937}
\definecolor{BoxHCU}{HTML}{0057C8}
\tcbset{hcubox/.style={enhanced, breakable, colback=white, boxrule=0.8pt, arc=2pt,
  left=4pt, right=4pt, top=2pt, bottom=2pt, toptitle=1pt, bottomtitle=1pt,
  fonttitle=\sffamily\bfseries\footnotesize, coltitle=white,
  before skip=6pt, after skip=6pt}}
\newtcolorbox{sourcebox}[1]{hcubox, colframe=BoxHCU, colbacktitle=BoxHCU, title={#1}}
\newtcolorbox{recordbox}[1]{hcubox, colframe=BoxInk, colbacktitle=BoxInk, title={#1}}
\definecolor{PanelAX}{HTML}{004488}
\definecolor{PanelGUI}{HTML}{7A5500}
\tcbset{hcupanel/.style={enhanced, colback=white, boxrule=0.7pt, arc=1.5pt,
  left=3pt, right=3pt, top=1pt, bottom=1pt, toptitle=0.5pt, bottomtitle=0.5pt,
  fonttitle=\sffamily\bfseries\scriptsize, coltitle=white,
  before skip=0pt, after skip=3pt}}
\newtcolorbox{axpanel}[1]{hcupanel, colframe=PanelAX, colbacktitle=PanelAX, title={#1}}
\newtcolorbox{guipanel}[1]{hcupanel, colframe=PanelGUI, colbacktitle=PanelGUI, title={#1}}
\newtcolorbox{plainpanel}[1]{hcupanel, colframe=BoxInk, colbacktitle=BoxInk, title={#1}}
\definecolor{PanelMuted}{HTML}{374151}

\fvset{breaklines, breakanywhere, fontsize=\scriptsize, fontfamily=zi4}
\RecustomVerbatimEnvironment{verbatim}{Verbatim}{
  fontsize=\small, frame=single, rulecolor=\color{black!25},
  framerule=0.4pt, framesep=8pt}
\usepackage{newunicodechar}
\newunicodechar{→}{\ensuremath{\rightarrow}}

\hypersetup{linkcolor=msftblue, citecolor=msftblue, urlcolor=msftblue,
  pdftitle={Hybrid Computer Use: A Unified Programmatic, Semantic, and Visual Interface for General Agents}}
\techreportshorttitle{Hybrid Computer Use}
\newcommand{\reportdate}{October, 2026}
\renewenvironment{abstract}{\begin{msfttitlebox}\setlength{\parindent}{0pt}\small\color{msftdark}}{%
  \par\vspace{0.14cm}%
  {\setlength{\parskip}{0.06cm}\small
  {\msftmetalabel{Project Page}\href{https://microsoft.github.io/Hybrid-Computer-Use}{https://microsoft.github.io/Hybrid-Computer-Use}\par}%
  {\msftmetalabel{Code}\href{https://github.com/microsoft/Hybrid-Computer-Use}{https://github.com/microsoft/Hybrid-Computer-Use}\par}%
  }%
  \end{msfttitlebox}\suppressfloats[t]}
\newcommand{\hcubibstyle}{unsrtnat}
\begin{document}
\thispagestyle{empty}
\noindent
\begin{minipage}[c]{0.5\linewidth}
\raggedright
\raisebox{-0.5\height}{\msftbrandmark}
\end{minipage}
\begin{minipage}[c]{0.49\linewidth}
\raggedleft
{\msftdatefont\small\color{msftgray}\reportdate}
\end{minipage}\par
\vspace{0.35em}
\noindent{\color{msftline}\rule{\linewidth}{0.8pt}\par}
\vspace{1.0em}
\begin{center}
{\msfttitlefont\fontsize{21}{25}\selectfont\color{msftdark}
Hybrid Computer Use:\\ A Unified Programmatic, Semantic, and Visual\\ Interface for General Agents\par}
\vspace{1.25em}
{\normalsize\rmfamily\color{msftdark}
Haoqing~Wang$^{1*}$, Xiaoyi~Zhang$^{1\dagger}$, Sicheng~Fan$^{2}$, Zhexin~Zhang$^{3}$, Xinjie~Zhang$^{1}$, Jiahao~Li$^{1}$, Bin~Li$^{1}$, Yan~Lu$^{1}$\par
\vspace{0.5em}
{\small\color{msftgray}$^{1}$Microsoft Research\quad $^{2}$Fudan University\quad $^{3}$Tsinghua University\par
}}
\end{center}
{\renewcommand{\thefootnote}{}\renewcommand{\footnotemark}{}%
\footnotetext{$^{*}$Work done during an internship at Microsoft Research Asia.}%
\footnotetext{$^{\dagger}$Project Lead (xiaoyizhang@microsoft.com).}}
\vspace{0.45em}
\newcommand{\channelresulttable}{%
\begin{wraptable}[9]{r}{0.6\textwidth}
\vspace{-10pt}
\caption{GPT-5.4 ablations, OSWorld 1.0. Darker red: larger losses vs.\ HCU. Runs: Appendix~\ref{app:ablation-details}.}
\label{tab:channels}
\centering\scriptsize\setlength{\tabcolsep}{3pt}
\begin{tabular}{lrrrrr}
\toprule
Configuration & Score (\%) $\uparrow$ & Steps $\downarrow$ & Input (M) $\downarrow$ & Time (s) $\downarrow$ & Cost (\$) $\downarrow$\\
\midrule
\rowcolor{HCUBlue!12}\textbf{HCU} & 65.59 & 28.60 & 0.941 & 351.3 & 2.52\\
w/o GUI channel & 66.08 & \cellcolor{WorseRed!10}32.16 & 0.909 & \cellcolor{WorseRed!6}352.4 & 2.45\\
w/o AX channel & \cellcolor{WorseRed!17}62.79 & \cellcolor{WorseRed!11}33.28 & \cellcolor{WorseRed!10}1.051 & \cellcolor{WorseRed!11}401.8 & \cellcolor{WorseRed!10}2.82\\
w/o skill & \cellcolor{WorseRed!12}64.00 & \cellcolor{WorseRed!11}33.47 & \cellcolor{WorseRed!6}0.962 & \cellcolor{WorseRed!9}385.5 & \cellcolor{WorseRed!6}2.57\\
\midrule
CUA Driver & \cellcolor{WorseRed!19}62.30 & \cellcolor{WorseRed!20}40.74 & \cellcolor{WorseRed!40}2.181 & \cellcolor{WorseRed!20}500.8 & \cellcolor{WorseRed!40}5.65\\
\bottomrule
\end{tabular}
\vspace{-6pt}
\end{wraptable}
}
\newcommand{\interfacethreeroundtable}{%
\begin{table}[!htbp]

\caption{Per-run Score (\%); mean $\pm$ sample standard deviation across three runs.}
\label{tab:interface-three-rounds}
\centering\small\setlength{\tabcolsep}{4pt}
\begin{tabular}{lrrrr}
\toprule
Configuration & Run 1 & Run 2 & Run 3 & Mean $\pm$ SD\\
\midrule
HCU & 66.16 & 65.65 & 64.95 & 65.59 $\pm$ 0.61\\[3pt]
HCU w/o GUI & 65.64 & 64.70 & 67.92 & 66.08 $\pm$ 1.66\\[3pt]
HCU w/o AX & 63.53 & 62.30 & 62.55 & 62.79 $\pm$ 0.65\\[3pt]
HCU w/o skill & 62.84 & 65.33 & 63.83 & 64.00 $\pm$ 1.25\\[3pt]
CUA Driver & 63.29 & 61.09 & 62.51 & 62.30 $\pm$ 1.12\\[3pt]
\bottomrule
\end{tabular}
\end{table}
}

\newcommand{\threeroundskilltable}{%
\begin{table}[!htbp]

\caption{Skill diagnostics pooled across three runs per configuration (2,166 traces; 361 tasks). Cells report event or task-run counts.}
\label{tab:ablation}
\centering\small\setlength{\tabcolsep}{4pt}
\begin{tabular}{lrr}
\toprule
Metric (numerator / denominator) & HCU w/o skill & HCU\\
\midrule
Schema rejection / calls & 2460/14160 & 147/10783\\[3pt]
Task-runs with schema rejection & 601/1083 & 132/1083\\[3pt]
Task-runs with consecutive schema failures & 484/1083 & 4/1083\\[3pt]
All AX/GUI errors / calls & 2686/14160 & 638/10783\\[3pt]
Stale-category errors / calls & 107/14160 & 88/10783\\[3pt]
Execution or assertion errors / calls & 66/14160 & 380/10783\\[3pt]
Other call errors / calls & 53/14160 & 23/10783\\[3pt]
Multi-action AX / AX action calls & 233/4241 & 495/3496\\[3pt]
Task-runs using multi-action AX & 117/1083 & 230/1083\\[3pt]
Multi-action GUI / GUI calls & 793/4915 & 342/2379\\[3pt]
Task-runs using AX \texttt{find} & 312/1083 & 388/1083\\[3pt]
Task-runs using \texttt{assert\_ax} & 2/1083 & 290/1083\\[3pt]
Task-runs requesting postconditions & 9/1083 & 173/1083\\[3pt]
\bottomrule
\end{tabular}
\end{table}
}
\newcommand{\skillroundtable}{%
\begin{table}[!htbp]

\caption{Per-run skill-use rates (\%). Definitions match Table~\ref{tab:ablation}.}
\label{tab:skill-rounds}
\centering\small\setlength{\tabcolsep}{4pt}
\begin{tabular}{llrr}
\toprule
Run & Metric & HCU w/o skill & HCU\\
\midrule
First & Schema rejection & 17.17 & 1.43\\[3pt]
First & Consecutive errors & 43.77 & 0.83\\[3pt]
First & Multi-action AX & 4.49 & 15.00\\[3pt]
Second & Schema rejection & 17.86 & 1.04\\[3pt]
Second & Consecutive errors & 45.71 & 0.00\\[3pt]
Second & Multi-action AX & 5.87 & 12.92\\[3pt]
Third & Schema rejection & 17.08 & 1.62\\[3pt]
Third & Consecutive errors & 44.60 & 0.28\\[3pt]
Third & Multi-action AX & 6.02 & 14.60\\[3pt]
\bottomrule
\end{tabular}
\end{table}
}

\newcommand{\guiswapmatrix}{%
\begin{table}[!htbp]

\caption{GUI action-space replacement within HCU: Score (\%); differences are Claude-style minus GPT-style.}
\label{tab:gui-swap-matrix}
\centering\small\setlength{\tabcolsep}{4pt}
\begin{tabular}{lrrr}
\toprule
Model & GPT-style & Claude-style & Difference (pp)\\
\midrule
Claude Opus 5.5 & 71.70 & 72.51 & +0.82\\[3pt]
GPT-5.4 & 65.59 & 63.56 & -2.03\\[3pt]
\bottomrule
\end{tabular}
\end{table}
}

\newcommand{\vmainresulttable}{%
\begin{table}[!htbp]
\caption{OSWorld 2.0 main results. Bold: best. $^{*}$ OpenAI's offline-subset result; unranked. $^{\S}$ Pre-0808: StateAct uses 0624; Gemini/Kimi task counts unspecified. }
\label{tab:main}
\centering\footnotesize
\setlength{\tabcolsep}{2.2pt}
\renewcommand{\arraystretch}{1.0}
\begin{tabular*}{\linewidth}{@{\extracolsep{\fill}}llrrrr@{}}
\toprule
System & Model & Partial (\%) $\uparrow$ & Binary (\%) $\uparrow$ & Out./1k $\downarrow$ & Cost (\$) $\downarrow$ \\
\midrule
Computer-use loop & \textcolor{gray}{GPT-6 Astra$^{*}$} & \textcolor{gray}{72.60} & \textcolor{gray}{--} & \textcolor{gray}{--} & \textcolor{gray}{--} \\
 & Claude Opus 5 & 68.31 & 31.43 & 102.87 & 28.43 \\
 & Muse Spark 1.3 & 66.90 & 32.00 & -- & -- \\
 & GPT-5.6 Sol & 62.72 & 27.34 & 36.14 & 13.94 \\
 & Gemini 3.8 Flash$^{\S}$ & 59.00 & -- & -- & -- \\
 & Kimi K3$^{\S}$ & 58.30 & -- & -- & -- \\
StateAct & Claude Opus 4.8$^{\S}$ & 61.60 & 26.90 & $\sim$100.00 & \textbf{$\sim$7.80} \\
\addlinespace[2pt]
CUA Driver (Codex) & GPT-6 Astra & 68.82 & 31.50 & 25.8 & 10.05 \\
\midrule
\textbf{HCU} (Codex) & GPT-6 Astra & \textbf{72.37} & 32.40 & \textbf{25.4} & 10.03 \\
\textbf{HCU} (Claude Code) & Claude Opus 5 & 72.06 & \textbf{35.19} & 118.29 & 22.67 \\
\bottomrule
\end{tabular*}
\end{table}
}

\begin{abstract}
Computer-use agents increasingly tackle tasks ranging from everyday desktop activities to long-horizon professional workflows.
Yet interfaces centered on screenshots and mouse-and-keyboard actions can impose a human-interaction prior, requiring repeated screen interpretation while leaving the computer's semantic and programmatic capabilities underexposed.
We argue for exposing these capabilities and letting models choose and combine them.
We introduce Hybrid Computer Use (HCU), a unified interface for general agents that combines three complementary channels: programmatic interaction through code and command-line tools, semantic interaction through accessibility (AX), and visual interaction through graphical user interfaces (GUI).
Our HCU supports efficient AX interaction and guards against stale-state actions during channel switching. A supplementary HCU skill guides agents in selecting and coordinating channels to use the unified interface effectively.
Our designed AX tools return compact observations, execute actions in batches, and report their effects as AX diffs, so the model needs fewer round trips.
On OSWorld 2.0, HCU achieves the highest Partial and Binary scores among the compared systems.
Analyzing agent trajectories, we find that all evaluated models improve with HCU yet allocate work across the channels differently and consistently outperform compared baselines in terms of both efficiency and task success, highlighting the value of unified computer-use interfaces that allow flexible channel selection and combination.
\end{abstract}

\begin{figure}[!htbp]
\centering
\begin{minipage}[c]{0.48\linewidth}
\centering
\includegraphics[width=\linewidth]{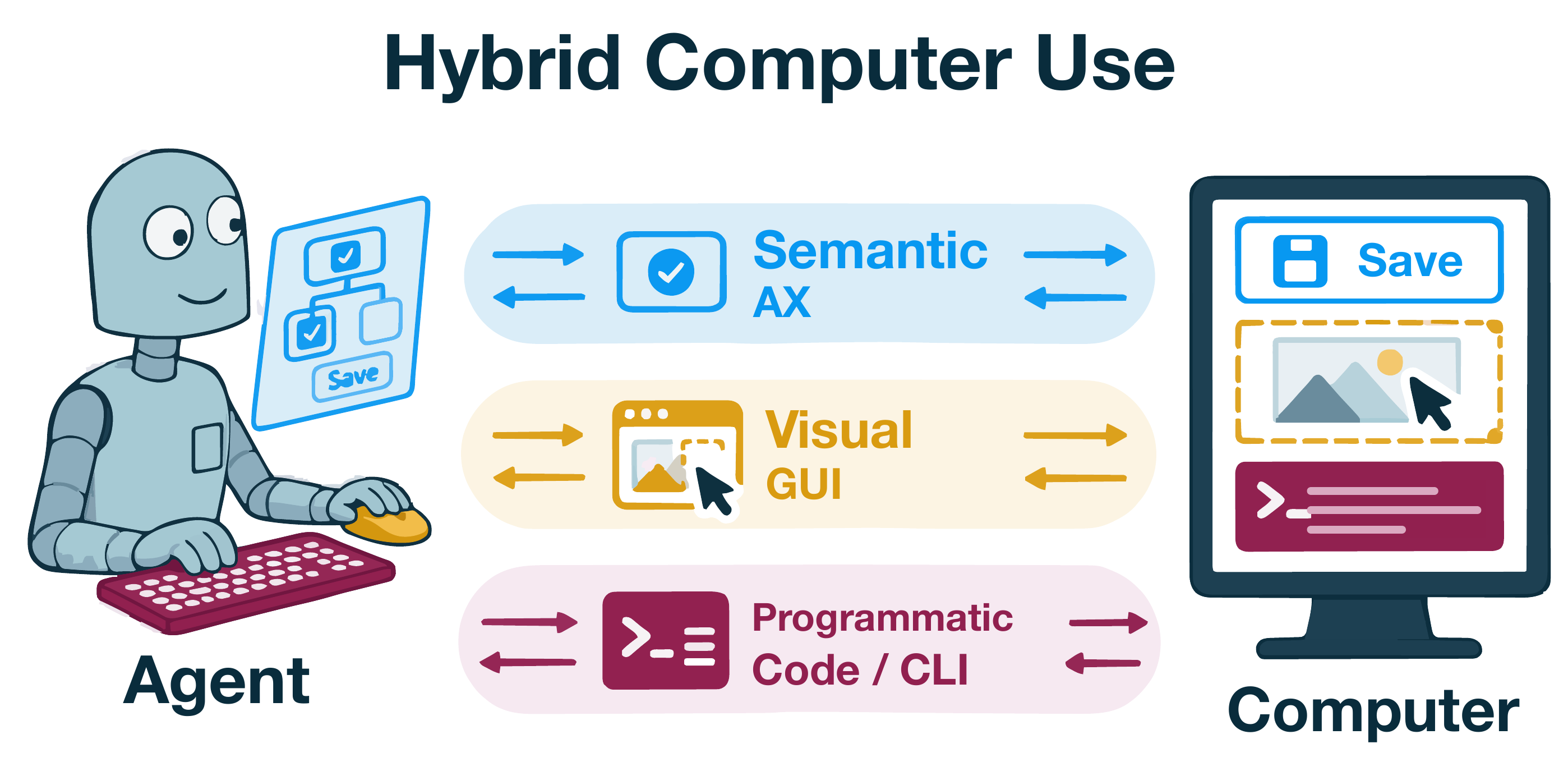}
\end{minipage}\hfill
\begin{minipage}[c]{0.50\linewidth}
\centering
\includegraphics[width=\linewidth]{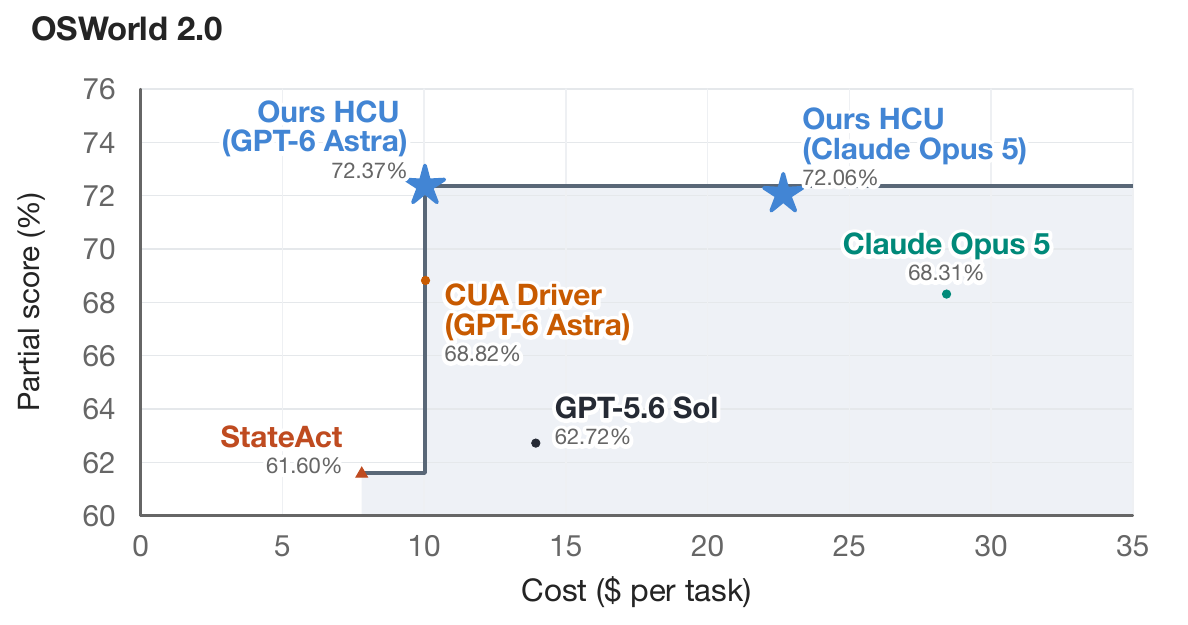}
\end{minipage}
\caption{Left: HCU's programmatic, semantic, and visual channels. Right: OSWorld 2.0 results. Stars mark HCU and the staircase marks the Pareto frontier, detailed in Section~\ref{sec:results}.
}
\label{fig:architecture}
\end{figure}

\section{Introduction}
\label{sec:intro}
AI agents are increasingly able to carry out user requests by operating software applications on computers directly. Computer-use research has expanded from everyday desktop tasks in OSWorld 1.0 \citep{xie2024osworld} to long-horizon workflows and specialized professional software in OSWorld 2.0 \citep{yuan2026osworld2}. 
As tasks become more involved, agents must sustain progress across many operations, making interaction efficiency important alongside task performance.

Most existing computer-use interfaces mirror how humans operate software: inspect the screen, locate a UI control, and use the mouse or keyboard. While this approach offers broad coverage with human-like perception and action, it also imposes a strong human-interaction prior on the action space of a frontier Large Language Model (LLM), overlooking more efficient channels for interacting directly with the system.
An agent with existing computer-use interface may spend substantial effort precisely locating a field by screen coordinates even when the application exposes its identity and value through accessibility data, or manipulate data through repeated click-and-drag actions when code could process it directly.

Modern computer systems offer a wide range of interaction tools that fall into three complementary channels: visual interaction through GUIs, structured semantic access through accessibility APIs, and direct programmatic access through code and command-line tools such as Bash.
Accessibility (AX), originally developed for assistive technologies and now also used for automation, exposes application UI controls, their properties, and supported operations \citep{qt2026accessibility,microsoft2025uiapatterns}. Code and command-line tools provide direct access to files, data, and software, while GUI interaction supplies visual evidence and control where structured access is insufficient.

Prior work \citep{sun2026ale,li2026weavebench} combines programmatic and visual interaction but makes limited use of the semantic channel provided by AX.
UFO2 combines UI Automation with GUI and application-API execution \citep{zhang2025ufo2}, while Tactile exposes semantic actions alongside OCR and visual fallbacks \citep{liu2026tactile}.
CUA Driver, an open-source computer-use interface, supports all three channels but favors AX-based interaction in the background to preserve the user's active desktop \citep{cua2026background}.
We build on these efforts to organize reference-addressed AX actions, in-call wait steps, and post-action feedback for efficient, sustained interaction, while exposing all three channels within a general agent for model-directed selection and combination.

We introduce Hybrid Computer Use (HCU), a unified interface for general agents that combines three complementary channels: programmatic, semantic, and visual interaction (Figure~\ref{fig:architecture}).
HCU adds AX and GUI tools to the agent's own programmatic tools via the Model Context Protocol (MCP), with cross-channel action guards to maintain a consistent view of the current desktop state and an HCU skill to guide the model in selecting and combining the exposed channels.
Our designed AX channel lets the model read a compact AX view, act on several targets in one call, and see the resulting changes without a separate observation.
Because all three channels act on the same desktop, an action in one can invalidate the evidence held by another; action guards therefore re-check targets and layouts against live state before each input. A supplementary HCU skill explains how to use these tools and when each channel is most effective, helping the model follow the execution and feedback contracts and reduce invocation errors.

We evaluate HCU on OSWorld 2.0 and OSWorld 1.0, and further evaluate it on WeaveBench, MacAgentBench, and Windows Agent Arena in Appendix~\ref{app:release-eval}. On OSWorld 2.0, HCU achieves the highest Partial and Binary scores among compared systems, advancing the performance--efficiency frontier (Figure~\ref{fig:architecture}, right).
With \mbox{GPT-6 Astra}, it reaches 72.37\% Partial, outperforming CUA Driver by 3.55 points. With Claude Opus~5, it improves Partial by 3.75 points and Binary by 3.76 points over the model's official leaderboard result, reaching 35.19\% Binary at 20\% lower Cost. On OSWorld 1.0, HCU outperforms CUA Driver across four other backbone models, each in its own harness, averaging 6.48 more points in Score, 17\% fewer Steps, and 30\% lower Cost, and reaches 70.13\% Score with GPT-5.4 while transferring these performance and efficiency gains to WeaveBench, MacAgentBench, and Windows Agent Arena at roughly half the Cost.
Component ablations and tool-call analyses reveal distinct channel preferences across models, yet gains consistently concentrate on tasks that use HCU's desktop tools. Within tasks, we observe that channels often play complementary roles, with screenshots supporting observation and AX enabling execution. The skill helps agents use the interface more reliably, reducing the proportion of AX and GUI calls rejected for malformed arguments from 17.37\% to 1.36\%.

Our contributions are threefold:
\begin{list}{\arabic{enumi}.}{%
  \usecounter{enumi}%
  \setlength{\leftmargin}{0pt}%
  \setlength{\itemindent}{1.5em}%
  \setlength{\labelwidth}{1em}%
  \setlength{\labelsep}{0.5em}%
  \setlength{\topsep}{4pt}%
  \setlength{\partopsep}{1pt}%
  \setlength{\itemsep}{2pt}%
  \setlength{\parsep}{2pt}%
}
  \item We introduce HCU, a unified interface for general agents that combines programmatic, semantic, and visual interaction. HCU pairs model-directed channel selection with cross-channel action guards and AX tools for compact observation, batched actions with wait steps, and post-action feedback.
  \item We show that HCU achieves the highest Partial and Binary scores on OSWorld 2.0 and advances the performance--efficiency frontier, and that it consistently improves both task performance and efficiency over CUA Driver across diverse backbone models, agent harnesses, and desktop platforms.
  \item We analyze the ablations and the cross-model comparison at the level of tool calls, showing that models given the same channels divide them differently, and we distill design lessons for future computer-use interfaces.
\end{list}

\section{Related Work}
\label{sec:related}

\paragraph{From GUI-based to hybrid computer use.}
Most computer-use agents observe screenshots and act through mouse-and-keyboard input. Work in this line improves grounding and planning, for example by routing GUI localization to specialist models and refining plans hierarchically \citep{agashe2025agents2}, and most results reported on OSWorld 1.0 and OSWorld 2.0 come from such screenshot-based loops \citep{xie2024osworld,yuan2026osworld2,osworld2026results}. Hybrid systems add programmatic access beside this visual loop. Some build dedicated multi-agent systems: CoAct-1 uses an orchestrator to delegate subtasks to a Programmer and a GUI Operator \citep{song2026coact}, and StateAct prioritizes direct program-state access and delegates visual interaction to a GUI subagent \citep{yang2026stateact}. Others place GUI interaction inside a general agent: the GUI-as-Tool setup in ALE and the desktop plugin evaluated in WeaveBench expose screenshots and mouse-and-keyboard actions as tools alongside the agent's programmatic tools \citep{sun2026ale,li2026weavebench}. While these works are training-free designs built around off-the-shelf models, Mage-VL-Cowork explores hybrid computer use from the training side \citep{mage2026cowork}. More broadly, SWE-agent shows how agent-computer interface design affects agent behavior and performance in software engineering \citep{yang2024sweagent}. Like ALE and WeaveBench, HCU extends a general agent without model training and leaves the division of work among channels to the model, while adding a semantic channel beside the programmatic and visual ones.

\paragraph{Semantic interaction with computers.}
Interfaces that expose semantic information differ in how they identify targets and execute actions. The original OSWorld 1.0 set-of-marks baseline uses AX information to identify targets, but maps element identifiers to coordinates for GUI input \citep{xie2024osworld,osworld2024reference}. UFO2 combines UI Automation and visual control detection with GUI and application-API execution \citep{zhang2025ufo2}. Tactile exposes compact desktop candidates and semantic actions with OCR and visual fallbacks \citep{liu2026tactile}. CUA Driver, our baseline, supports element-addressed actions and coordinate input under a best-effort background policy \citep{cua2026background}. In browsers, Playwright MCP provides accessibility snapshots and explicit target references \citep{playwrightmcp2025}, while Cocoa-Agent combines DOM queries and element-addressed actions with screenshots and programmatic tools \citep{cocoabench2026}. HCU uses explicit target references on desktop AX and adds batched AX actions with wait steps in the same call, cross-channel action guards, and AX diffs as post-action feedback.

\paragraph{Benchmarks and efficiency.}
OSWorld 1.0 is a widely used benchmark for computer-use agents in real desktop environments \citep{xie2024osworld}, and OSWorld 2.0 extends this evaluation to long-horizon real-world workflows \citep{yuan2026osworld2}. Early evaluations centered on GUI interaction; recent benchmarks increasingly examine hybrid workflows and cross-platform desktop use \citep{li2026weavebench,fu2026macagentbench,bonatti2024windowsagentarena,bai2026recreationworld}.
Following cost-aware agent evaluation \citep{kapoor2024agents}, we report task performance together with resource use across OSWorld 2.0, OSWorld 1.0, and the cross-platform evaluations in Appendix~\ref{app:release-eval}.

\Needspace{8\baselineskip}
\section{Hybrid Computer Use Interface}
\label{sec:method}
\label{sec:overview}

\begin{figure}[!htb]
\centering
\includegraphics[width=\linewidth]{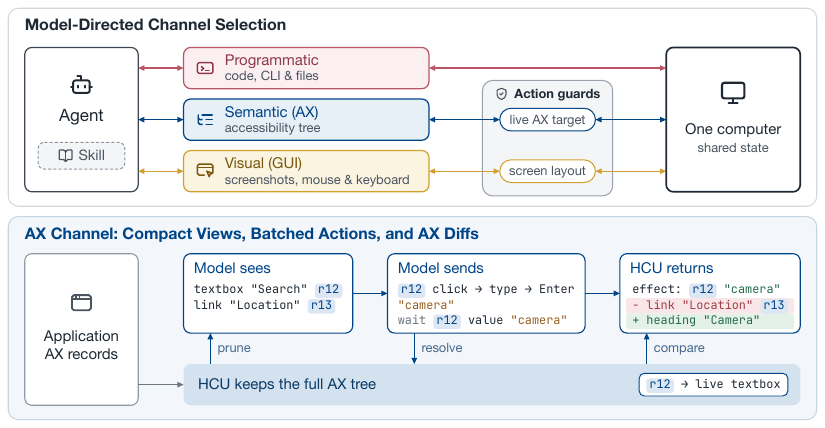}
\caption{HCU overview. Top: channels, skill, and action guards. Bottom: a compact AX view, batched actions with a wait step, and the returned effect summary and AX diff.}
\label{fig:interface-structure}
\end{figure}

HCU combines three interaction channels in one agent's task context: programmatic, semantic, and visual (Figure~\ref{fig:interface-structure}, top). The programmatic channel uses the agent's own code, shell, and file tools. HCU adds the semantic and visual channels through two desktop tools exposed via the Model Context Protocol (MCP) \citep{mcp2025architecture}: the AX channel reads application accessibility trees and executes reference-addressed actions, while the GUI channel takes screenshots and sends mouse and keyboard input (Appendix Table~\ref{tab:hcu-contract}). All three channels act on the same computer. Action guards check AX targets and GUI layouts against live state before input. A supplementary skill explains how to use the interface.

The agent uses HCU within its existing observation--action loop. Given the task and available observations, the model chooses a channel to inspect state or act; the selected tool executes the request and returns its result to the same context, from which the model decides whether to continue, inspect further, or change channels. Observation and execution need not use the same channel: a screenshot can supply visual context for a subsequent AX action. Channel selection remains model-directed. HCU adds no planner, subagents, or loop of its own, so the same interface attaches to different agent harnesses (Appendix~\ref{app:design-configuration}).

The following subsections describe the AX channel's observation, execution, and feedback design (Section~\ref{sec:verification}), the action guards that support switching between channels (Section~\ref{sec:validity}), and the skill that explains how to use the interface (Section~\ref{sec:guidance}).

\subsection{Accessibility for Computer Control}
\label{sec:verification}
\label{sec:feedback}
\label{sec:axnature}

AX exposes a structured representation that applications or UI frameworks supply for assistive technologies: a tree of elements with roles, names, values, states, and supported operations. It lets an agent identify and operate controls without locating every target in pixels, but using it to control a computer presents three challenges.

\begin{list}{\arabic{enumi}.}{%
  \usecounter{enumi}%
  \setlength{\leftmargin}{0pt}%
  \setlength{\itemindent}{1.5em}%
  \setlength{\labelwidth}{1em}%
  \setlength{\labelsep}{0.5em}%
  \setlength{\topsep}{4pt}%
  \setlength{\partopsep}{0pt}%
  \setlength{\itemsep}{2pt}%
  \setlength{\parsep}{0pt}%
}
\item \textbf{Coverage depends on what applications expose.} Applications choose which properties and operations to expose through AX. Custom-drawn content may have no accessible elements, accessible objects need not map one-to-one to widgets, and virtualized items may be absent until rendered \citep{qt2026accessibility,microsoft2025uiavirtualization}. AX therefore provides incomplete coverage of visible content and available interactions. CUA Driver addresses this by returning the tree and a screenshot together by default, so the model can use pixels when AX omits or misrepresents a target \citep{cua2026modalities}. HCU retains the acquired AX records, marks unavailable information, and lets the model request a screenshot when semantic evidence is insufficient.
\item \textbf{Changing targets.} Controls can share names, move, disappear, or be recreated; AT-SPI explicitly marks defunct objects \citep{gnome2026atspistates}. A label or tree position therefore cannot establish identity across actions. CUA Driver instructs the agent to refresh element indices before acting and rejects ambiguous window targets \citep{cua2026actionresult,cua2026mcpinstructions}. HCU binds target references independently of the displayed view and resolves them against live elements before each action, including actions within a batch.
\item \textbf{Verifying action effects.} AT-SPI reports invocation, which does not establish that the intended state change occurred \citep{gnome2026atspiaction}. CUA Driver verifies caller-defined postconditions through a separate operation \citep{cua2026actionresult}. HCU lets the agent compose actions and optional AX wait steps in one batch, such as selecting a checkbox and waiting for its checked state before continuing, without a separate verification call.
\end{list}

CUA Driver prioritizes best-effort background input to preserve the user's active desktop, with foreground input available on request \citep{cua2026background}. In contrast, HCU prioritizes completing the requested action, using native AX operations or focused input as needed. Its efficiency-oriented design also gives the agent more control over how work is composed: compact views reduce observation size, while batched actions, optional wait steps, and AX diffs let the agent combine execution, verification, and state feedback within a single call.

\subsubsection{Composable Semantic Interaction}
\label{sec:axdesign}

These properties motivate an AX interface that lets the agent compose actions while HCU validates targets, runs requested wait steps, and returns evidence for continuation. Figure~\ref{fig:interface-structure} (bottom) illustrates this interaction: a compact AX view supplies target references, a batch combines actions with optional wait steps, and post-action feedback supports the next decision.

\paragraph{Compact AX views and target references.}
HCU separates the AX records it retains from the compact AX view shown to the model. It removes structural wrappers and condenses repeated content in the view while retaining the acquired records for target identification and state comparison. Target references are bound to elements before this compression, so shortening the view does not change which element a reference identifies. The retained records also support AX diffs after an action, while \texttt{find} and scoped observations let the model retrieve targets omitted from the compact AX view. Appendix~\ref{app:observations} gives same-capture examples and measures the reduction in tokens.

\paragraph{Batched actions and wait steps.}
HCU lets the agent compose known actions and optional AX wait steps in a single call. The \texttt{act} mode executes the steps in order within one application window, checking each target with the AX target guard before input (Section~\ref{sec:validity}). A wait step sends no input and holds until an expected state appears, such as a checkbox becoming checked or a dialog opening, before later steps run; placed last, it verifies the final state. This combines batching, also used for GUI interaction \citep{li2026weavebench}, with state verification akin to Playwright assertions \citep{playwright2026assertions}. If an action fails, a wait times out, or window continuity is lost, HCU stops the batch and reports the stopping point and execution evidence. This lets the agent decide how to continue without blindly repeating earlier actions.

\paragraph{Post-action feedback.}
HCU returns per-target effect summaries and an AX diff by default after an action call, showing changed values, window changes, and target references relative to the last eligible observation delivered to the model. The agent can instead request a screenshot or execution evidence only, according to what it needs for its next decision. Returning this feedback with the action lets the agent continue without a separate observation when the information suffices (traces and samples in Appendix Figure~\ref{fig:interaction} and ~\ref{app:feedback-samples}).

HCU reports only what the available evidence supports, so the agent is not led to treat input delivery as a verified effect. Like CUA Driver \citep{cua2026actionresult}, it distinguishes known non-delivery from an unknown effect after possible input. The agent uses this distinction and the returned continuation instructions to decide whether to retry, inspect state, or stop.

\Needspace{14\baselineskip}
\subsection{Guarding Actions Across Channels}
\label{sec:validity}

\begin{wrapfigure}{r}{0.56\textwidth}
\vspace{-12pt}
\begin{minipage}{\linewidth}
\setlength{\intextsep}{0pt}
\begin{algorithm}[H]
\caption{Cross-channel action guards (simplified)}
\label{alg:channelguards}
\footnotesize
\algrenewcommand{\algorithmicindent}{0.9em}
\algrenewcommand{\algorithmicrequire}{\textbf{Input:}}
\algtext*{EndFor}
\algtext*{EndIf}
\makeatletter
\patchcmd{\ALG@doentity}{\item[]\nointerlineskip}{\item[]\vskip-\prevdepth\nointerlineskip}{}{\PackageError{HCU}{Algorithm end-line spacing patch failed}{}}
\makeatother
\begin{algorithmic}[1]
\Require \begin{tabular}[t]{@{}l@{}}
a targeted AX batch or GUI coordinate batch $A$\\
last screenshot $F$, if any
\end{tabular}
\For{each $a \in A$ in order}
  \If{$a$ is an AX action} \Comment{AX target guard}
    \State $t \gets \Call{LiveTarget}{a.\mathrm{ref}}$ \Comment{$\bot$ if identity differs}
    \State $\mathit{ok} \gets t \neq \bot \wedge \Call{CanExecute}{a, t}$
  \Else \Comment{GUI layout guard}
    \State $\mathit{ok} \gets F \neq \bot \wedge \textproc{LiveLayout}() = F.\mathrm{layout}$
  \EndIf
  \If{not $\mathit{ok}$} \Comment{no input for $a$ or later actions}
    \State \Return fresh evidence or a request to observe
  \EndIf
  \State \Call{Execute}{$a$}
\EndFor
\end{algorithmic}
\end{algorithm}
\end{minipage}
\vspace{-8pt}
\end{wrapfigure}

The model observes the desktop through discrete tool outputs rather than a continuous view of its state. After an action, it may not know which earlier observations remain valid, especially when combining channels. An AX action may open a dialog over coordinates from an earlier screenshot, or a GUI action may replace a control addressed by an AX reference. Reusing that evidence can send input to an unintended target without the model noticing. HCU therefore leaves channel choice to the model while using action guards to check the targets and layouts on which its actions depend immediately before input (Algorithm~\ref{alg:channelguards}).

The AX target guard resolves each target reference immediately before execution and checks the element's identity and action-specific requirements. If the target is missing, ambiguous, or replaced, or an action-specific requirement is not met, HCU rejects the action without sending input. It does not redirect the action to a similarly named element. The GUI layout guard checks whether the layout underlying the model's chosen coordinates still matches the last screenshot. Before coordinate input, it compares the current desktop bounds, visible top-level windows, stacking order, active window, and target-point window layout with those associated with that screenshot. If the layout has changed, HCU blocks the affected input and returns a fresh screenshot so the model can choose new coordinates.

Both guards inspect current state regardless of which channel changed it, including earlier actions in the same channel. Detected mismatches lead to explicit rejection before input, allowing the model to revise its next action (Appendix~\ref{app:state-failures}; recorded rejection in Figure~\ref{fig:guard}).

\subsection{Skill for Effective Interface Use}
\label{sec:guidance}

HCU provides a supplementary skill that explains how to choose and compose the interface's capabilities. We distilled its instructions from recurring mistakes in development trajectories, including malformed arguments and misused target references, following the trajectory-informed refinement of usage instructions in SWE-agent \citep{yang2024sweagent}. The skill addresses three aspects of interaction: (1) \textbf{Channel capabilities.} The skill explains what programmatic, semantic, and visual interaction can provide, including their observation scope and limitations, so the model can decide which channels to use and combine for the task. (2) \textbf{Efficient interaction.} It explains how to use target references, batch known actions with wait steps, and continue from returned feedback, requesting additional observations only when needed. (3) \textbf{Verification and recovery.} It distinguishes input delivery, local UI changes, and task outcomes; explains how to verify saved or applied results; and describes how to recover from stale targets, invocation errors, or insufficient evidence.

The skill makes these instructions concrete through rules for common invocation and recovery mistakes, while the tool schemas define argument shapes. It contains no task-specific solutions; Appendix~\ref{app:skill-text} presents its text. The skill ablation in Section~\ref{sec:channels} shows reduced invocation errors and improved interaction efficiency; Appendix~\ref{app:versions} describes instruction integration.

\section{Experiments}
\label{sec:experiments}

\subsection{Experimental Setup}
\label{sec:experimentalsetup}

\paragraph{Benchmarks.}
The main experiment uses the OSWorld 2.0 full set (0808 release; \citealp{yuan2026osworld2}), and the ablations and cross-model study use OSWorld 1.0 \citep{xie2024osworld}; Appendix~\ref{sec:setup} lists the task sets. Appendix~\ref{app:release-eval} reports additional evaluations across OSWorld 1.0, WeaveBench \citep{li2026weavebench}, MacAgentBench \citep{fu2026macagentbench}, and Windows Agent Arena \citep{bonatti2024windowsagentarena}.

\paragraph{Models and comparisons.}
Each model always runs in the same harness: GPT-6 Astra and GPT-5.4 in Codex, Claude Opus 5 and Claude Opus 5.5 in Claude Code, Gemini 3.8 Flash in Gemini CLI, and Grok 4.7 in Grok Build. We therefore name configurations by interface and model. In HCU--CUA Driver comparisons, both interfaces use the same harness execution flow and programmatic tools within each model; auxiliary harness features are restricted to focus the comparison on the interface. The main experiment evaluates HCU (GPT-6 Astra) and HCU (Claude Opus 5) at each model's highest reasoning effort, with CUA Driver (GPT-6 Astra) as the matched baseline and reported results as context. The interface-component ablations use GPT-5.4 (xhigh); the cross-model comparison adds Claude Opus 5.5 (low), Gemini 3.8 Flash (high), and Grok 4.7 (xhigh). We gray out the official GPT-6 Astra results since they were evaluated internally on a subset of tasks. Appendix~\ref{app:configurations} details all configurations.

\paragraph{Metrics and budgets.}
We report Partial (mean task score) and Binary (full-credit rate) on OSWorld 2.0 and Score on OSWorld 1.0, along with per-task resource use. The main experiment uses a 500-step limit and the ablations a 100-step limit. We average results across runs within each configuration. Appendix~\ref{app:protocol} details the runs, protocols, and metric accounting.

\subsection{Main Results}
\label{sec:results}

Table~\ref{tab:main} compares HCU with results reported by the official OSWorld 2.0 website \citep{osworld2026results}, StateAct \citep{yang2026stateact}, and model vendors \citep{openai2026astra,meta2026muse13,google2026gemini38flash,moonshot2026kimik3} (Appendix~\ref{app:v2references}). Among the full-set results in Table~\ref{tab:main}, HCU achieves the highest Partial (72.37\% with GPT-6 Astra) and Binary (35.19\% with Claude Opus 5). Only StateAct costs less than HCU with GPT-6 Astra, about \$7.80 against \$10.03 per task, at 61.60\% Partial; the two form the performance--efficiency frontier (Figure~\ref{fig:architecture}, right).

\vmainresulttable

With Claude Opus 5, HCU scores 3.75 points higher in Partial and 3.76 points higher in Binary than the official leaderboard result, at 20.26\% lower Cost. With GPT-6 Astra, HCU outperforms the matched CUA Driver baseline by 3.55 points in Partial and 0.90 in Binary with 29.81\% fewer tool calls at nearly identical Cost (Appendix Table~\ref{tab:codexreported}).

\paragraph{Cross-platform evaluations.}
With GPT-5.4, the released version of HCU (Appendix~\ref{app:release-eval}) also outperforms CUA Driver on WeaveBench (53.25\% vs.\ 46.37\%), the MacAgentBench subset (25 vs.\ 24 of 28 tasks), and Windows Agent Arena (80.86\% vs.\ 71.89\%) at 46--51\% lower Cost per task.

\raggedbottom
\Needspace{16\baselineskip}
\subsection{Interface Components}
\label{sec:channels}

\channelresulttable
We remove the GUI channel, the AX channel, or the skill one at a time to assess each component's contribution (Table~\ref{tab:channels}).

\paragraph{HCU beats the baseline on every measure.}
HCU scores 3.29 points higher than the baseline while using 29.80\% fewer Steps, 56.85\% fewer input tokens, 29.85\% less Time, and 55.36\% lower Cost. Every component contributes to this efficiency: each one cuts Steps by 11--15\%, so the full interface needs the fewest Steps of all configurations.

\begin{wrapfigure}[9]{r}{0.32\textwidth}
\vspace{-14pt}
\centering
\includegraphics[width=\linewidth]{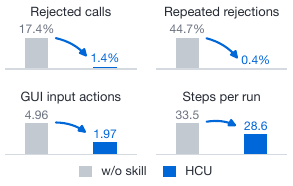}
\vspace{-16pt}
\caption{Skill effects.}
\label{fig:skill-inline}
\vspace{-10pt}
\end{wrapfigure}
\paragraph{AX contributes most.}
AX is the channel HCU adds to GUI-based computer use: it exposes the applications' control tree as named, reference-addressed targets and returns observed state changes after action batches. Added to the harness's programmatic tools and the GUI channel, it raises Score by 2.79 points and cuts Steps by 14.08\%, input tokens by 10.44\%, Time by 12.58\%, and Cost by 10.64\%, the largest gain of any component in Score, input, Time, and Cost. The GUI channel complements AX primarily through interaction efficiency: removing GUI leaves Score within run-to-run variation while increasing Steps by 12.45\% (Appendix~\ref{app:ablation-details}). AX thus drives the Score gain and most of the resource savings, while GUI saves Steps when visual context or pixel input is required; Section~\ref{sec:channeluse} shows how the two divide observation and execution.

\paragraph{Execution-centered AX design supports efficient interaction.}
HCU's AX channel integrates guarded execution, optional wait steps, and post-action state in one call (Section~\ref{sec:axdesign}). With only AX and programmatic tools, HCU w/o GUI exceeds CUA Driver by 3.79 points in Score while using 21.06\% fewer Steps, 58.33\% fewer input tokens, 29.63\% less Time, and 56.70\% lower Cost. CUA Driver also offers element-addressed actions and separates delivery from effect \citep{cua2026actionresult}, but leaves verification and renewed observation to separate operations. Both interfaces can sequence actions programmatically; HCU builds verification and feedback into the AX call itself.

\paragraph{The skill improves calling reliability and interaction efficiency.}
With the skill (Section~\ref{sec:guidance}), the share of AX and GUI calls rejected for malformed arguments falls from 17.37\% to 1.36\%, and the share of task-runs that repeat such rejections on the same tool while varying the arguments falls from 44.69\% to 0.37\% (Figure~\ref{fig:skill-inline}). Feedback selection also changes: 87\% of AX action calls omit the \texttt{observe} field, selecting the default AX diff, compared with 0.2\% without the skill. GUI input actions decrease from 4.96 to 1.97 per task. These behavioral changes accompany a 14.55\% reduction in Steps and a 1.59-point gain in Score (Appendix~\ref{app:metrics}).

\subsection{Agent Behavior Across Models}
\label{sec:modelstudy}

\begin{figure}[!b]
\centering
\includegraphics[width=0.90\linewidth]{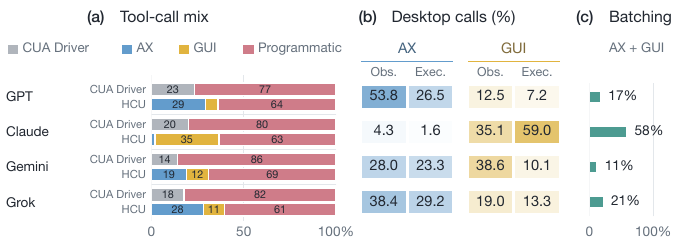}
\caption{Interface use. (a) Mean per-task tool-call shares (Programmatic: the agent's programmatic tools; CUA Driver includes browser tools). (b) Within-HCU AX/GUI shares: observation (AX reads, GUI screenshots) versus execution (AX actions, GUI inputs). (c) AX/GUI acting calls with at least two actions (\%).}
\label{fig:crossmodel-use}
\end{figure}

We compare HCU with CUA Driver across four models (Table~\ref{tab:crossmodel}), examining whether the gains transfer and how each model obtains them. Here GPT, Claude, Gemini, and Grok denote GPT-5.4, Claude Opus 5.5, Gemini 3.8 Flash, and Grok 4.7, respectively.

\Needspace{10\baselineskip}
\begin{wraptable}[10]{r}{0.58\textwidth}
\vspace{-10pt}
\caption{HCU versus CUA Driver on OSWorld 1.0. Resource changes are relative to CUA Driver; darker blue indicates larger improvements. Details: Appendix~\ref{app:crossmodel}.}
\label{tab:crossmodel}
\centering\scriptsize\setlength{\tabcolsep}{3pt}
\begin{tabular}{lrrrrrrr}
\toprule
& \multicolumn{3}{c}{Score (\%) $\uparrow$} & \multicolumn{4}{c}{Change with HCU (\%) $\downarrow$}\\
\cmidrule(lr){2-4}\cmidrule(lr){5-8}
Model & CUA Driver & HCU & $\Delta$ (pp) & Steps & Input & Time & Cost\\
\midrule
GPT-5.4 & 62.30 & 65.59 & \cellcolor{HCUBlue!8}+3.29 & \cellcolor{HCUBlue!16}$-$30 & \cellcolor{HCUBlue!30}$-$57 & \cellcolor{HCUBlue!16}$-$30 & \cellcolor{HCUBlue!29}$-$55\\
Claude Opus 5.5 & 68.68 & 71.62 & \cellcolor{HCUBlue!7}+2.94 & \cellcolor{HCUBlue!1}$-$1 & \cellcolor{HCUBlue!17}$-$32 & +6 & \cellcolor{HCUBlue!4}$-$8\\
Gemini 3.8 Flash & 54.67 & 67.57 & \cellcolor{HCUBlue!30}+12.90 & \cellcolor{HCUBlue!9}$-$17 & \cellcolor{HCUBlue!20}$-$39 & \cellcolor{HCUBlue!15}$-$29 & \cellcolor{HCUBlue!23}$-$44\\
Grok 4.7 & 67.90 & 74.71 & \cellcolor{HCUBlue!16}+6.81 & \cellcolor{HCUBlue!11}$-$22 & \cellcolor{HCUBlue!9}$-$18 & \cellcolor{HCUBlue!7}$-$13 & \cellcolor{HCUBlue!7}$-$13\\
\bottomrule
\end{tabular}
\vspace{-6pt}
\end{wraptable}

\paragraph{HCU improves every model in Score, input, and Cost.}
Every model scores higher with HCU than with CUA Driver, by 2.94 to 12.90 points, and uses 18--57\% fewer input tokens at 8--55\% lower Cost (Table~\ref{tab:crossmodel}). Steps and Time also decrease for GPT, Gemini, and Grok. For Claude, Steps remain nearly unchanged while Time increases by 6\%; we examine the contrasting patterns of desktop use behind its Step count below.

\paragraph{Models use HCU's channels differently.}
\label{sec:channeluse}
All four models allocate a larger share of tool calls to HCU's AX and GUI desktop tools than to CUA Driver (Figure~\ref{fig:crossmodel-use}a). Among HCU's AX and GUI calls, AX accounts for 80\% for GPT, 68\% for Grok, and 51\% for Gemini. Claude instead uses GUI for 94\% of these calls (Figure~\ref{fig:crossmodel-use}b) and batches actions more often: 58\% of its acting calls contain two or more actions, compared with 11--21\% for the other models (Figure~\ref{fig:crossmodel-use}c).

\paragraph{Models combine observation and execution across channels.}
GPT, Gemini, and Grok combine visual observation with semantic action within the same task. On 29--31\% of the tasks where each model used GUI, all GUI calls were screenshots, while the model executed actions through AX (Appendix~\ref{app:channel-use}). In one recorded task, a screenshot confirmed an entered value missing from AX feedback, after which the agent continued acting through AX (Appendix~\ref{app:findings}). By leaving channel choice to the model, HCU allows observation and execution to use different channels, enabling the model to draw on their complementary capabilities as the task requires.

\Needspace{24\baselineskip}
\begin{wrapfigure}{r}{0.49\textwidth}
\vspace{-8pt}
\centering
\includegraphics[width=\linewidth]{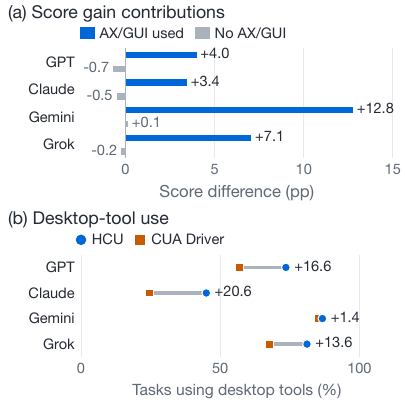}
\caption{(a) Contributions to the HCU--CUA Driver Score difference, grouped by HCU AX/GUI use. (b) Tasks using each interface's desktop tools; labels show HCU--CUA Driver differences in points.}
\label{fig:crossmodel-gains}
\vspace{-8pt}
\end{wrapfigure}
\paragraph{Score gains concentrate on tasks with AX or GUI use.}
Across all four models, nearly all of the overall Score gain over CUA Driver comes from tasks where HCU makes at least one AX or GUI call (Figure~\ref{fig:crossmodel-gains}a).
We group tasks by whether HCU makes at least one such call in each run and average the groups' weighted contributions across runs. The group without these calls contributes only $-0.7$ to +0.1 points to the overall Score difference (Appendix~\ref{app:gain-decomposition}).

\paragraph{Overall Steps mask differences in interaction efficiency.}
On tasks where both configurations invoke their desktop tools (AX or GUI), HCU uses 19--40\% fewer Steps across the four models. HCU also invokes desktop tools on more tasks than CUA Driver (Figure~\ref{fig:crossmodel-gains}b). For Claude, tasks where only HCU invokes desktop tools gain 7.6 points in Score, accompanied by 53\% more Steps. This group's weighted increase largely offsets the savings where both configurations use desktop tools, explaining Claude's nearly unchanged overall Steps despite its higher Score. Appendix~\ref{app:gain-decomposition} provides the complete decomposition.

\section{Discussion and Conclusion}
\label{sec:discussion}

HCU exposes programmatic, semantic, and visual interaction to a general agent and leaves their selection and combination to the model. It achieves the highest Partial and Binary scores on OSWorld 2.0, and consistently outperforms CUA Driver in both task performance and efficiency across diverse backbone models, agent harnesses, and desktop platforms.
These results suggest that interfaces built on how people operate software, inspecting the screen and using the mouse and keyboard, cost agents both performance and efficiency. Given AX, three of the four models send most of their desktop calls through it (Section~\ref{sec:channeluse}). This echoes the bitter lesson \citep{sutton2019bitter}: a strong human-like interface that fixes how the computer is used may limit what increasingly capable models can do with it. The capabilities HCU exposes already exist on standard desktops and require no model retraining.

Finally, the AX channel exposes element states that code can read and act on. This makes it possible to treat the desktop as a programmable environment and to bring desktop work into code execution and programmatic tool calling \citep{wang2024codeact}. HCU's batched actions with wait steps (Section~\ref{sec:axdesign}) provide a direct substrate for this composition. We hope this work encourages interfaces that fully leverage what computers already offer to release model capabilities on computer use.

\label{main-end}

\pdfbookmark[1]{References}{references}
\bibliography{references}
\bibliographystyle{\hcubibstyle}

\clearpage
\appendix
\hcuappendixtitle
Appendix~\ref{app:release-eval} reports additional evaluations across Linux (OSWorld 1.0 and WeaveBench), macOS (MacAgentBench), and Windows (Windows Agent Arena). Appendix~\ref{app:protocol} specifies the experimental setup and OSWorld 2.0 comparisons, Appendix~\ref{app:components} analyzes the interface-component ablations, and Appendix~\ref{app:trajectories} examines agent behavior across models. Finally, Appendix~\ref{app:design} details HCU's architecture and implementation, and Appendix~\ref{app:tools} reproduces the tool definitions, base policy, and hybrid skill.

\section{Cross-Platform and Additional Benchmark Evaluation}
\label{app:release-eval}

\begin{figure}[!htbp]
\centering
\includegraphics[width=\linewidth]{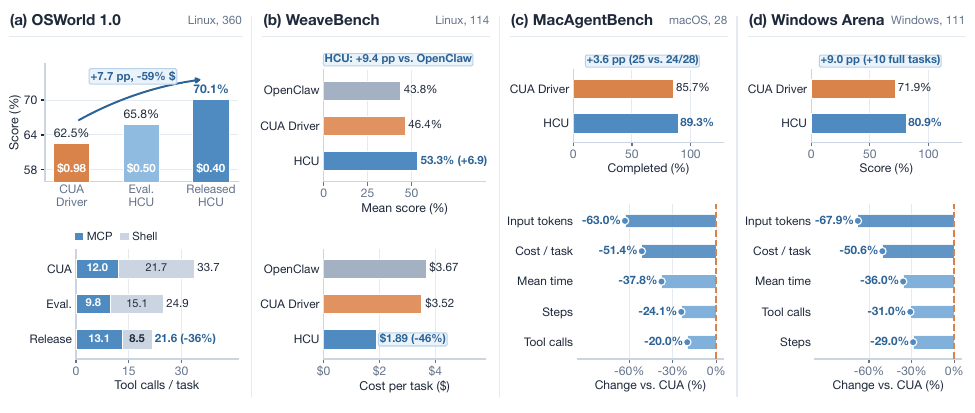}
\caption{Evaluation of HCU across benchmarks and platforms (GPT-5.4). (a) OSWorld 1.0 (Linux, 360 tasks): Score, Cost per task, and tool-call shift from Shell to desktop tools across CUA Driver, evaluated HCU, and released HCU. (b) WeaveBench (Linux, 114 tasks): mean score and Cost per task across Official OpenClaw, CUA Driver, and HCU. (c) MacAgentBench UI-dependent subset (macOS, 28 tasks) and (d) Windows Agent Arena (Windows, 111 tasks): task performance and resource reductions relative to CUA Driver.}
\label{fig:release-cross-platform}
\end{figure}

The experiments in the main text use the evaluated version of HCU (Appendix~\ref{app:versions}). Following those experiments, we refined the interface and skill into the released version described in Appendices~\ref{app:design} and~\ref{app:tools}. This appendix evaluates HCU across additional benchmarks and operating systems (Figure~\ref{fig:release-cross-platform}): OSWorld 1.0 \citep{xie2024osworld} and WeaveBench \citep{li2026weavebench} on Linux, a UI-dependent subset of MacAgentBench \citep{fu2026macagentbench} on macOS, and Windows Agent Arena \citep{bonatti2024windowsagentarena} on Windows.

\subsection{OSWorld 1.0}
\label{app:release-osworld}

\paragraph{Setup.}
Released HCU runs with GPT-5.4 at xhigh reasoning effort in Codex 0.146.0 under the base policy and hybrid skill in Appendix~\ref{app:tools}. The task set, the 100-step and 1,800-second limits per task, evaluator, and task images match the OSWorld 1.0 ablations. We compare released HCU against the three-run means of evaluated HCU and CUA Driver from the ablations across the 360 evaluated tasks.

\begin{table}[!htbp]
\caption{CUA Driver, evaluated HCU, and released HCU on OSWorld 1.0 with GPT-5.4 (xhigh) in Codex, 360 tasks. CUA Driver and Evaluated HCU: mean of three runs. Values are per task unless noted; tool calls count desktop calls (AX and GUI) and shell commands (programmatic channel).}
\label{tab:release-osworld}
\centering\small
\begin{tabular}{@{}lrrrrr@{}}
\toprule
Metric & CUA Driver & Evaluated HCU & Released HCU & vs.\ Evaluated & vs.\ CUA Driver\\
\midrule
Score (\%) & 62.47 & 65.77 & 70.13 & +4.36 pp & +7.66 pp\\
Full-score tasks (\%) & 60.00 & 62.41 & 66.67 & +4.26 pp & +6.67 pp\\
Steps & 40.58 & 28.45 & 24.92 & $-$12.4\% & $-$38.6\%\\
Tool calls & 33.67 & 24.94 & 21.63 & $-$13.3\% & $-$35.8\%\\
\quad Desktop calls & 12.02 & 9.83 & 13.10 & +33.2\% & +9.0\%\\
\quad Shell commands & 21.66 & 15.11 & 8.52 & $-$43.6\% & $-$60.7\%\\
Time (s) & 499 & 350 & 296 & $-$15.3\% & $-$40.7\%\\
Input tokens (K) & 2,174.5 & 927.1 & 717.7 & $-$22.6\% & $-$67.0\%\\
Output tokens & 12,992 & 11,205 & 9,471 & $-$15.5\% & $-$27.1\%\\
Cost (\$) & 0.983 & 0.504 & 0.402 & $-$20.2\% & $-$59.1\%\\
\bottomrule
\end{tabular}
\end{table}

\paragraph{Results.}
Relative to evaluated HCU, released HCU raises Score by 4.36 points and the share of full-score tasks by 4.26 points while using 12.4\% fewer Steps, 22.6\% fewer input tokens, 15.3\% less Time, and 20.2\% lower Cost per task (Table~\ref{tab:release-osworld}). Relative to CUA Driver on the same 360 tasks, released HCU raises Score by 7.66 points and cuts Steps by 38.6\%, Time by 40.7\%, input tokens by 67.0\%, and Cost by 59.1\%. It also shifts work from shell commands to the desktop tools: desktop calls rise to 13.10 per task while shell commands fall to 8.52 ($-43.6\%$ vs.\ evaluated HCU and $-60.7\%$ vs.\ CUA Driver), in line with the server instructions directing all desktop input through the AX and GUI tools (Appendix~\ref{app:tool-definitions}).

On the 189 tasks that released HCU and all three evaluated HCU runs complete with full score, Steps per task fall from 22.35 to 19.57, Time by 19.3\%, and Cost by 19.2\%.

\subsection{WeaveBench}
\label{app:release-weavebench}

\paragraph{Setup.}
We also evaluate HCU on all 114 tasks of WeaveBench \citep{li2026weavebench}, which tests long-horizon hybrid workflows across data analysis, design, documents, desktop systems, games, operations, professional software, and the web. HCU and CUA Driver both run with GPT-5.4 at medium reasoning effort in Codex 0.146.0, and we compare them with the official GPT-5.4 (medium) OpenClaw baseline. All three arms are scored with the benchmark's OpenClaw agent-as-judge rubric. Tool calls count recorded invocations, with one batched call counted once.

\begin{table}[!htbp]
\caption{Three-arm comparison on all 114 WeaveBench tasks with GPT-5.4 (medium). Values are per task unless noted.}
\label{tab:release-weavebench}
\centering\small
\begin{tabular}{@{}lrrr@{}}
\toprule
Metric & Official OpenClaw & HCU & CUA Driver\\
\midrule
Mean score (\%) & 43.84 & 53.25 & 46.37\\
Tasks with score $\ge 0.8$ & 23/114 & 24/114 & 12/114\\
Total tool calls & 9,458 & 9,465 & 11,494\\
Tool calls & 82.96 & 83.03 & 100.82\\
Input tokens & 3,537,196 & 4,478,251 & 8,712,900\\
Output tokens & 32,677 & 26,821 & 31,671\\
Cost (\$) & 3.672 & 1.889 & 3.518\\
Time (min) & 21.19 & 32.59 & 40.76\\
\bottomrule
\end{tabular}
\end{table}

\paragraph{Results.}
Across all 114 tasks, HCU reaches a mean score of 53.25\%, exceeding CUA Driver by 6.88 points and the official OpenClaw baseline by 9.41 points (Table~\ref{tab:release-weavebench}). HCU also doubles the count of tasks with score at least 0.8 relative to CUA Driver (24 vs.\ 12) while using 17.6\% fewer tool calls, 48.6\% fewer input tokens, 20.0\% less Time, and 46.3\% lower Cost per task (\$1.889 vs.\ \$3.518), and cuts Cost per task by 48.6\% relative to the official OpenClaw baseline (\$1.889 vs.\ \$3.672).

\subsection{MacAgentBench UI-Dependent Subset (macOS)}
\label{app:release-macagentbench}

\paragraph{Setup.}
To compare the native macOS backends of HCU and CUA Driver directly, we also evaluate both interfaces on a 28-task subset of MacAgentBench \citep{fu2026macagentbench}. Many native macOS applications in MacAgentBench expose AppleScript dictionaries or command-line utilities that allow an agent with Bash access to complete the task without interacting with the graphical interface, leaving no distinction between desktop-tool interfaces. We therefore select 28 tasks across Clock, Safari, System Settings, and Contacts that require operating the graphical interface to complete, providing a direct comparison between HCU and CUA Driver on macOS. Both arms run with GPT-5.4 at xhigh reasoning effort in Codex, with Bash enabled and 100-step and 1,800-second limits per task.

\begin{table}[!htbp]
\caption{HCU and CUA Driver on the 28 UI-dependent macOS tasks of MacAgentBench with GPT-5.4 (xhigh). Values are per task unless noted.}
\label{tab:release-macagentbench}
\centering\small
\begin{tabular}{@{}lrrr@{}}
\toprule
Metric & CUA Driver & HCU & Change\\
\midrule
Completed tasks & 24/28 (85.7\%) & 25/28 (89.3\%) & +3.6 pp\\
Steps & 52.21 & 39.61 & $-$24.1\%\\
Tool calls & 44.54 & 35.64 & $-$20.0\%\\
Time, mean (s) & 876.0 & 544.5 & $-$37.8\%\\
Time, median (s) & 659.5 & 419.3 & $-$36.4\%\\
Input tokens & 3,637,053 & 1,344,550 & $-$63.0\%\\
Output tokens & 13,500 & 11,174 & $-$17.2\%\\
Cost (\$) & 1.4899 & 0.7245 & $-$51.4\%\\
Total Cost, 28 tasks (\$) & 41.7185 & 20.2865 & $-$51.4\%\\
\bottomrule
\end{tabular}
\end{table}

\paragraph{Results.}
On the 28 UI-dependent macOS tasks, HCU completes 25 of 28 tasks (89.3\%) compared with 24 of 28 (85.7\%) for CUA Driver, while substantially reducing interaction cost across every efficiency metric (Table~\ref{tab:release-macagentbench}). Because these tasks are relatively straightforward for GPT-5.4 (xhigh) under the 100-step limit, the limit rarely binds and both interfaces reach high completion rates; the advantage of HCU therefore manifests primarily in interaction efficiency, consistent with the below-limit regime in Appendix~\ref{app:termination}. HCU cuts mean Time by 37.8\% (544.5\,s vs.\ 876.0\,s; median $-36.4\%$), Steps by 24.1\% (39.61 vs.\ 52.21), input tokens by 63.0\% (1.34M vs.\ 3.64M), output tokens by 17.2\%, and Cost by 51.4\% (\$0.7245 vs.\ \$1.4899 per task; \$20.29 vs.\ \$41.72 total). Across HCU's 998 tool calls on these tasks, AX accounts for 49.8\% (497), Shell for 28.6\% (285), and GUI for 21.6\% (216), showing that the semantic channel carries half of all tool calls on macOS when tasks require desktop UI interaction.

\subsection{Windows Agent Arena (Windows)}
\label{app:release-waa}

\paragraph{Setup.}
To evaluate the native Windows backend (UIA3 and Win32) under the same public AX and GUI contract, we compare HCU and CUA Driver on 111 tasks of Windows Agent Arena (WAA; \citealp{bonatti2024windowsagentarena}) with GPT-5.4 in Codex. One task that did not complete under CUA Driver due to a Calc deadlock and out-of-memory error triggered by the driver is scored as 0 under the fixed 111-task denominator.

\begin{table}[!htbp]
\caption{HCU and CUA Driver on 111 Windows Agent Arena tasks with GPT-5.4 in Codex. Values are per task unless noted.}
\label{tab:release-waa}
\centering\small
\begin{tabular}{@{}lrrr@{}}
\toprule
Metric & CUA Driver & HCU & Change\\
\midrule
Score (\%) & 71.89 (79.80/111) & 80.86 (89.75/111) & +8.97 pp\\
Full / partial / zero tasks & 78 / 2 / 31 & 88 / 2 / 21 & +10 full\\
Steps, mean (median) & 24.45 (20.5) & 17.35 (14.0) & $-$29.0\%\\
Tool calls, mean (median) & 23.81 (20.0) & 16.43 (13.0) & $-$31.0\%\\
Time, mean (s) & 344.3 & 220.3 & $-$36.0\%\\
Time, median (s) & 199.5 & 171.8 & $-$13.9\%\\
Input tokens, total & 155,591,157 & 49,975,768 & $-$67.9\%\\
Output tokens, total & 927,272 & 697,839 & $-$24.7\%\\
Cost (\$) & 0.664 & 0.325 & $-$51.0\%\\
Total Cost, all tasks (\$) & 73.00 & 36.06 & $-$50.6\%\\
\bottomrule
\end{tabular}
\end{table}

\paragraph{Results.}
On Windows Agent Arena, HCU reaches 80.86\% Score (89.75/111) compared with 71.89\% (79.80/111) for CUA Driver, gaining 8.97 points and solving 10 additional tasks with full score (88 vs.\ 78; Table~\ref{tab:release-waa}). On the 110 tasks where both arms complete evaluation, HCU still leads by 8.14 points (80.68\% vs.\ 72.54\%, winning on 14 tasks against 6 for CUA Driver, with 74 shared full scores and 16 shared zeros). Across the benchmark, HCU cuts mean Time by 36.0\% (220.3\,s vs.\ 344.3\,s; median 171.8\,s vs.\ 199.5\,s), Steps by 29.0\% (17.35 vs.\ 24.45), total input tokens by 67.9\% (50.00M vs.\ 155.59M), output tokens by 24.7\% (697.8K vs.\ 927.3K), and total Cost by 50.6\% (\$36.06 vs.\ \$73.00; \$0.325 vs.\ \$0.664 per task).

\section{Experimental Setup and OSWorld 2.0 Details}
\label{app:protocol}
\label{app:results}

This appendix specifies the task sets, agent configurations, metrics, and OSWorld 2.0 comparison details for Section~\ref{sec:experiments}.

\subsection{Benchmarks, Configurations, and Metrics}
\label{sec:setup}
\label{app:configurations}
\label{app:records}
\label{app:metrics}
\label{app:trajectory-data}
\label{app:case-archives}
\label{app:case-index}

\paragraph{Task sets.}
The main experiment uses the OSWorld 2.0 full set (108 tasks; 0808 release). The OSWorld 1.0 comparisons cover 361 tasks (excluding eight Google Drive tasks): 93 multi-application tasks, 47 each in LibreOffice Calc and Impress, 46 in Chrome, 26 in GIMP, 24 operating-system tasks, 23 each in VS Code and LibreOffice Writer, 17 in VLC, and 15 in Thunderbird.

\paragraph{Models, harnesses, and tool exposure.}
GPT-6 Astra and GPT-5.4 run in Codex; Claude Opus 5 and Claude Opus 5.5 run in Claude Code; Gemini 3.8 Flash runs in Gemini CLI; and Grok 4.7 runs in Grok Build. The main experiment evaluates GPT-6 Astra and Claude Opus 5 at each model's highest reasoning effort under a 500-step limit. The OSWorld 1.0 comparisons use GPT-5.4 (xhigh), Claude Opus 5.5 (low), Gemini 3.8 Flash (high), and Grok 4.7 (xhigh) under 100-step and 1,800-second limits per task on isolated Linux desktops. Within each harness, auxiliary features (web search, memory, subagents, and planning modes) are disabled so that both HCU and CUA Driver share the same execution loop, base policy, and programmatic tools (Bash and file tools), with foreground desktop input enabled on all calls. Table~\ref{tab:capabilities} summarizes the exposed components across the four HCU ablation configurations.

\capabilitytable

\paragraph{Metrics and aggregation.}
For task scores $s_i \in [0,1]$, OSWorld 1.0 Score and OSWorld 2.0 Partial denote the mean score $Q_{\mathrm{mean}}=\frac{100}{N}\sum_{i=1}^{N}s_i$, and OSWorld 2.0 Binary denotes the full-credit rate $\frac{100}{N}\sum_{i=1}^{N}\mathbf{1}[s_i=1]$. Efficiency is measured by delivered model responses (``Steps''), recorded tool calls, input and output tokens, agent execution Time, and Cost at provider list prices. One model response may issue multiple tool calls, and one AX or GUI call may execute a batch of steps. For configurations with multiple runs (three runs for all five GPT-5.4 ablation configurations, totaling 5,415 task-runs, and two runs for Claude Opus 5.5 HCU), metrics are averaged within each task before averaging across tasks; relative changes are ratios of per-task means.

\paragraph{Evaluated and released versions.}
\label{app:versions}
The experiments in Section~\ref{sec:experiments} use the evaluated version of HCU, which shares the same channels, target references, action guards, batching, and post-action diffs as the released version. In the released version (Appendices~\ref{app:design} and~\ref{app:tools}), in-batch checks are unified as ordered \texttt{wait} steps (replacing action- and batch-level postconditions and \texttt{assert\_ax}), window discovery uses \texttt{apps} (replacing \texttt{observation\_target}), the batch limit increases from 32 to 64 steps, and every AX and GUI result starts with a JSON evidence line carrying a public error code. The released version also adds GUI \texttt{zoom}, AX \texttt{select}, \texttt{settle} waits, the per-target effect report, and per-target-point layout checks on Linux. Appendix~\ref{app:release-eval} evaluates the released version across OSWorld 1.0, WeaveBench, MacAgentBench, and Windows Agent Arena.

\subsection{OSWorld 2.0 Results and Baselines}
\label{app:v2details}
\label{app:v2matched}
\label{app:v2references}
\label{app:frontier}

\paragraph{HCU runs and matched CUA Driver comparison.}
In the matched GPT-6 Astra comparison in Codex with Code Mode enabled (Table~\ref{tab:codexreported}), HCU improves Partial by 3.55 points (72.37\% vs.\ 68.82\%) and Binary by 0.90 points (32.4\% vs.\ 31.5\%), while using 29.81\% fewer tool calls (78.4 vs.\ 111.7), 9.57\% fewer Steps (85 vs.\ 94), and 21.00\% less mean Time (38.25 vs.\ 48.42 minutes; median 32.95 vs.\ 36.80 minutes) at virtually identical Cost (\$10.03 vs.\ \$10.05). With Claude Opus 5 in Claude Code, HCU reaches 72.06\% Partial and 35.19\% Binary, averaging 75.8 outer agent turns, 88.0 Steps, 81.6 tool calls, 8.16M input tokens, and 113.95 minutes per task at \$22.67, reducing Cost by 20.26\% relative to the official leaderboard result for Claude Opus 5 (\$28.43).

\begin{table}[!htbp]
\caption{HCU versus CUA Driver with GPT-6 Astra in Codex, OSWorld 2.0 full set (0808). Code Mode; 500-step limit.}
\label{tab:codexreported}
\centering\small\setlength{\tabcolsep}{12pt}
\begin{tabular}{@{}lrr@{}}\toprule
Metric & CUA Driver & HCU \\
\midrule
Partial (\%) & 68.82 & 72.37 \\
Binary (\%) & 31.5 & 32.4 \\
Steps & 94 & 85 \\
Tool calls & 111.7 & 78.4 \\
Input tokens / task (M) & 6.9 & 6.9 \\
Output tokens / task (k) & 25.8 & 25.4 \\
Time (min), mean & 48.42 & 38.25 \\
Time (min), median & 36.80 & 32.95 \\
Cost (\$) & 10.05 & 10.03 \\
\bottomrule\end{tabular}
\end{table}

\paragraph{Reported baselines and Pareto frontier.}
Table~\ref{tab:main} and Figure~\ref{fig:architecture} (right) include official OSWorld 2.0 leaderboard and chart records \citep{osworld2026results} at each model's highest reasoning effort under the 500-step limit, together with author- and vendor-reported results for StateAct \citep{yang2026stateact} (61.60\% Partial, 26.90\% Binary, $\sim$\$7.80 and $\sim$100k output tokens on release 0624), Muse Spark 1.3 \citep{meta2026muse13,meta2026muse13methodology} (66.90\% Partial, 32.00\% Binary on 0808), Gemini 3.8 Flash \citep{google2026gemini38flash,google2026gemini38methodology} (59.00\% Partial, pre-0808), Kimi K3 \citep{moonshot2026kimik3,alibaba2026qwen38max} (58.30\% Partial, pre-0808), and OpenAI's offline-subset report for GPT-6 Astra \citep{openai2026astra} (72.60\% Partial). In Figure~\ref{fig:architecture} (right), the Pareto frontier over all full-set records with paired Cost is formed by StateAct and HCU (GPT-6 Astra).

\section{Interface Components}
\label{app:components}

This appendix complements Section~\ref{sec:channels} with per-run ablation results, skill error and batching statistics across all 2,166 HCU traces, and the GUI action-space replacement experiment.

\subsection{Channel and Skill Ablations Across Runs}
\label{app:ablation-details}

\interfacethreeroundtable

In every run, HCU scores above HCU w/o AX, HCU w/o skill, and CUA Driver, and requires the fewest Steps of all five configurations: 28.4--28.8 Steps per task, against more than 31 for every other arm (Table~\ref{tab:interface-three-rounds}). Across the three runs on the same 361 tasks, single-run Scores span 1.21 points for HCU, 1.23 for HCU w/o AX, 2.20 for CUA Driver, 2.49 for HCU w/o skill, and 3.22 for HCU w/o GUI. HCU leads HCU w/o AX by 2.40--3.35 points and CUA Driver by 2.44--4.56 points in every run, whereas the contrast with HCU w/o GUI changes sign across runs (+0.52 and +0.95 in runs 1--2, $-2.97$ in run 3) while GUI consistently cuts Steps by 12.45\%.

\subsection{Skill, Invocation Errors, and Batching}
\label{app:guidance}
\label{app:batch-feedback-use}

\threeroundskilltable

\begin{figure}[!htbp]
\centering
\includegraphics[width=\linewidth]{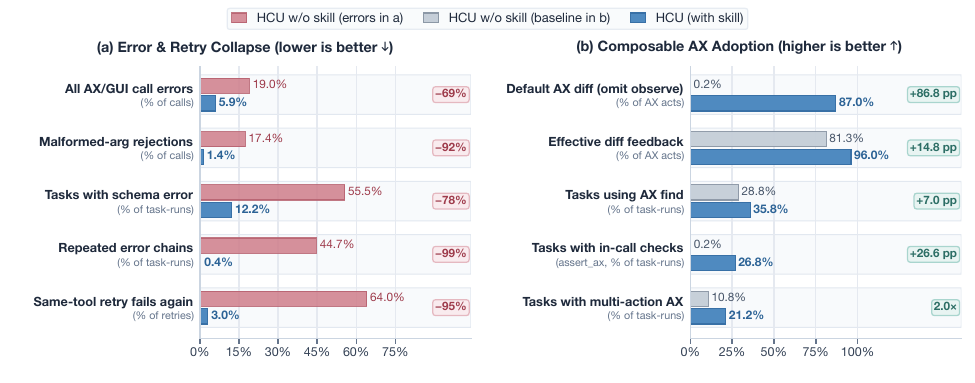}
\caption{Skill effects across 2,166 OSWorld 1.0 traces (GPT-5.4). (a) Collapse of invocation errors, schema-error chains, and repeated same-tool failures. (b) Adoption of composable AX features (default diff feedback, multi-action batches, \texttt{find}, and in-call checks).}
\label{fig:skill-error-details}
\end{figure}

Table~\ref{tab:ablation} and Figure~\ref{fig:skill-error-details} compare the 1,083 task-runs of HCU against the 1,083 task-runs of HCU w/o skill. Adding the skill reduces total AX and GUI call errors from 2,686/14,160 (18.97\%) to 638/10,783 (5.92

The skill also shifts how the model composes calls: multi-action AX batches rise from 233/4,241 to 495/3,496, targeted \texttt{find} use rises from 312 to 388 task-runs, in-call checks in the evaluated version (unified as \texttt{wait} steps in the released version; Appendix~\ref{app:versions}) rise from two to 290 (\texttt{assert\_ax}) and nine to 173 (postconditions), default-or-explicit diff selection rises from 81.25\% to 96.02\% of AX action calls, and explicit follow-up observations drop from 1,553 to 1,258 calls.

\subsection{GUI Action-Space Replacement}
\label{app:gui-replacement}

\guiswapmatrix

To test whether GUI action vocabulary affects interface use, we swap HCU's default GPT-style GUI vocabulary for a Claude-style vocabulary (Table~\ref{tab:gui-action-spaces}) while keeping AX, programmatic tools, batching, and layout guards fixed. Claude Opus 5.5 (low, two runs per arm: 71.29/72.10 vs.\ 72.16/72.86) and GPT-5.4 (xhigh, three GPT-style runs vs.\ two Claude-style runs: 63.20/63.91) each perform best when the GUI vocabulary matches their native calling conventions: Claude Opus 5.5 gains 0.82 points with the Claude-style vocabulary, while GPT-5.4 scores 2.03 points higher with the default GPT-style vocabulary (Table~\ref{tab:gui-swap-matrix}). Under its native vocabulary, convention errors per 1,000 AX/GUI calls drop from 92.49 (80/865, including 27 unsupported \texttt{triple\_click} calls) to 48.14 (44/914) for Claude Opus 5.5, and are 10.44 (38/3,639) versus 41.48 (156/3,761) for GPT-5.4.

\section{Agent Behavior Across Models}
\label{app:trajectories}

This appendix supports Section~\ref{sec:modelstudy} with per-application Score gains, cross-channel observation--execution patterns, the decomposition of Score and Steps by desktop-tool use, step-limit termination analysis, and post-error transitions.

\subsection{Application Breakdown and Cross-Channel Interaction}
\label{app:crossmodel}
\label{app:channel-use}
\label{app:findings}

\begin{figure}[!htbp]
\centering
\includegraphics[width=\linewidth]{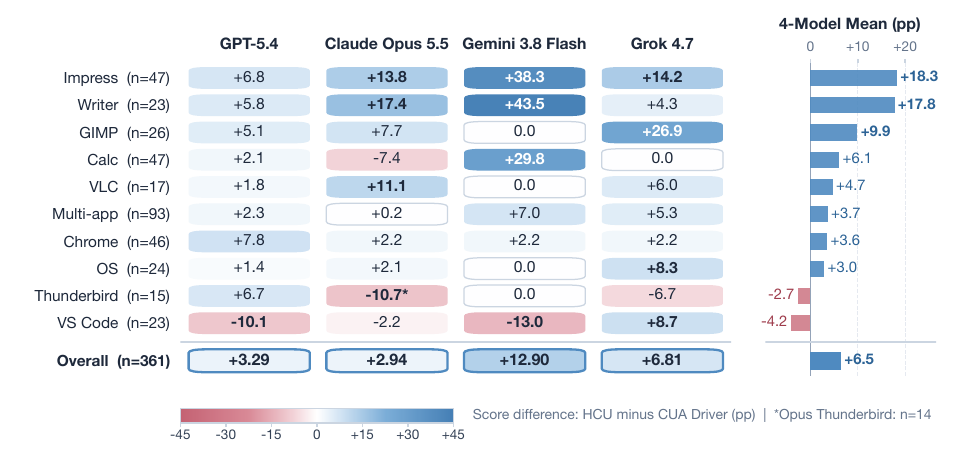}
\caption{HCU--CUA Driver Score differences by application; tasks are averaged over runs before pairing. $n$: paired tasks; asterisk: Claude Opus 5.5's 14-task Thunderbird comparison.}
\label{fig:crossmodel-apps}
\end{figure}

Every configuration in Table~\ref{tab:crossmodel} evaluates all 361 OSWorld 1.0 tasks except CUA Driver with Claude Opus 5.5, where the provider's cybersecurity safeguards refused one task (pairing the remaining 360 tasks). Per-run HCU Scores are 66.16, 65.65, and 64.95 for GPT-5.4 and 71.29 and 72.10 for Claude Opus 5.5. Figure~\ref{fig:crossmodel-apps} breaks down Score gains by application: Gemini 3.8 Flash gains most in LibreOffice Calc, Impress, and Writer (+30, +38, and +43 points), while Grok 4.7 gains most in GIMP and Impress (+27 and +14 points).

\begin{figure}[!htbp]
\centering
\includegraphics[width=\linewidth]{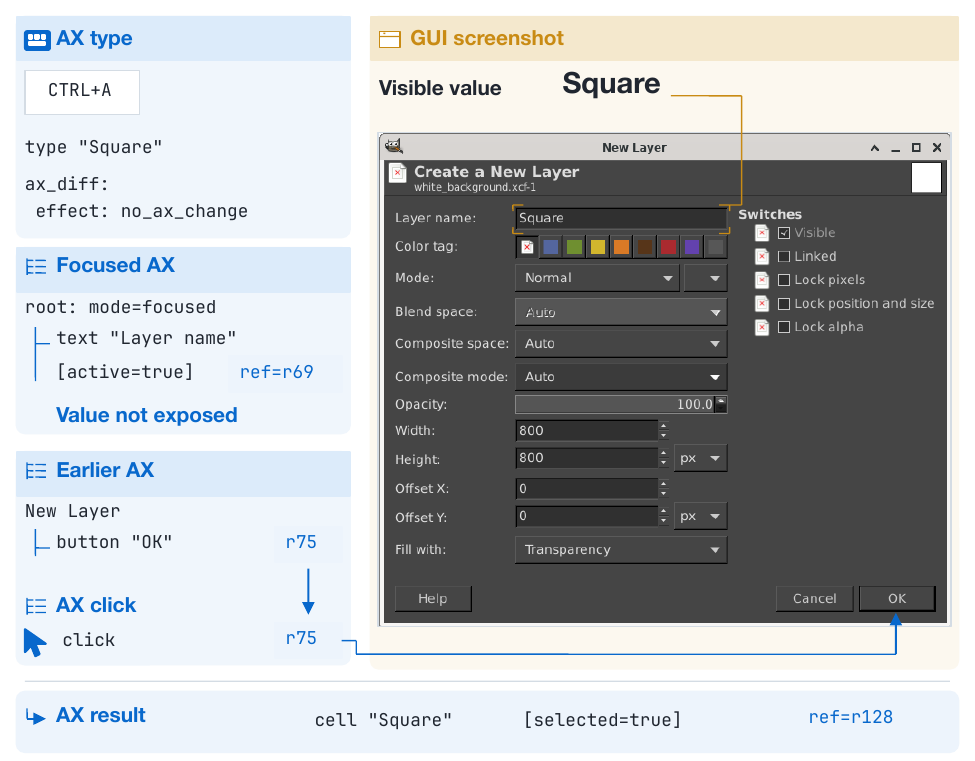}
\caption{Recorded GIMP interaction, HCU (GPT-5.4), OSWorld 1.0. A screenshot reveals the value omitted by AX; execution resumes through the earlier OK reference.}
\label{fig:gimp-cross-channel}
\end{figure}

\paragraph{Combining visual observation with semantic execution.}
Among tasks where HCU issued GUI calls, GUI was used exclusively for screenshots on 77/189 tasks for GPT-5.4, 134/275 for Gemini 3.8 Flash, 100/254 for Grok 4.7, and 28/158 for Claude Opus 5.5; on 59 (31\%), 81 (29\%), 77 (30\%), and 1 (0.6\%) of those tasks, respectively, the model executed desktop actions through AX. Figure~\ref{fig:gimp-cross-channel} shows a recorded GIMP trace: after AX types ``Square'', the AX tree omits the updated field value, a GUI screenshot confirms it visually, and execution resumes via the AX reference \texttt{r75} for OK.

\subsection{Decomposing Score Gains and Steps}
\label{app:gain-decomposition}
\label{app:input-accounting}
\label{tab:steps-desktop-decomposition}

\begin{figure}[!htbp]
\centering
\includegraphics[width=\linewidth]{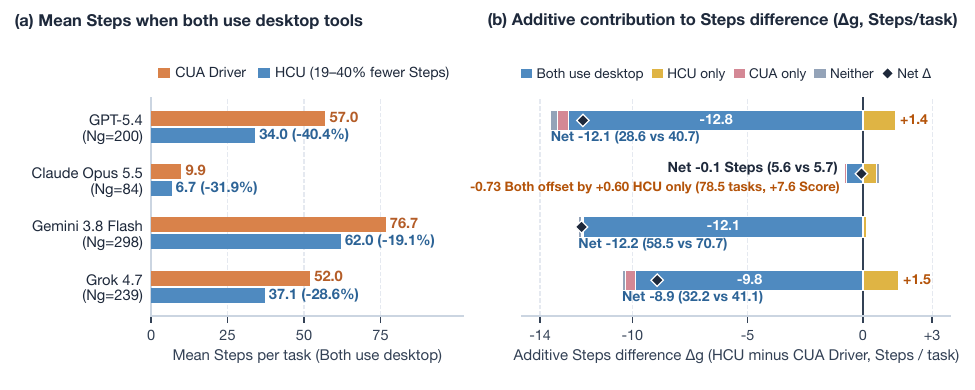}
\caption{Steps decomposition by desktop-tool use across models. (a) Mean Steps per task and equivalent task counts $N_g$ on tasks where both interfaces use desktop tools. (b) Additive contributions $\Delta_g$ to the overall Steps difference (negative favors HCU) and net per-task Steps.}
\label{fig:steps-decomposition}
\end{figure}

Figure~\ref{fig:crossmodel-gains}a decomposes the total Score difference $\Delta = \sum_g \Delta_g$ by whether HCU invoked AX or GUI on task $i$ in run $r$: $\Delta_g=\frac{100}{NR}\sum_{r=1}^{R}\sum_{i=1}^{N}\mathbf{1}[g_{ir}=g](s^{H}_{ir}-\bar{s}^{C}_i)$. Similarly, Figure~\ref{fig:steps-decomposition} decomposes the per-task Steps difference across four desktop-use groups (both interfaces use desktop tools, HCU only, CUA Driver only, or neither), weighting run pairs equally: $w_g=\frac{1}{NR_HR_C}\sum_{i,r,s}\mathbf{1}[g_{irs}=g]$ and $\Delta_g=w_g(\overline{t}^{H}_g-\overline{t}^{C}_g)$.

On tasks where both interfaces invoke desktop tools, HCU reduces Steps by 19.14\%--40.38\% across all four models (Figure~\ref{fig:steps-decomposition}). For Claude Opus 5.5, saving $-0.73$ Steps per task in the both-desktop group is offset by $+0.60$ Steps from the 78.5 tasks where only HCU invokes desktop tools (which gain +7.64 points), yielding the net $-0.07$ Step change ($-1.17\%$). Across applications, LibreOffice Calc, Impress, Writer, and Thunderbird (132/361 tasks) account for 69.4\%, 77.6\%, and 40.5\% of net Step savings relative to CUA Driver for GPT-5.4, Gemini 3.8 Flash, and Grok 4.7, and 70.0\% relative to HCU w/o AX for GPT-5.4.

\subsection{Budget Exhaustion and Post-Error Behavior}
\label{app:termination}
\label{app:invocation-errors}

\begin{figure}[!htbp]
\centering
\includegraphics[width=\linewidth]{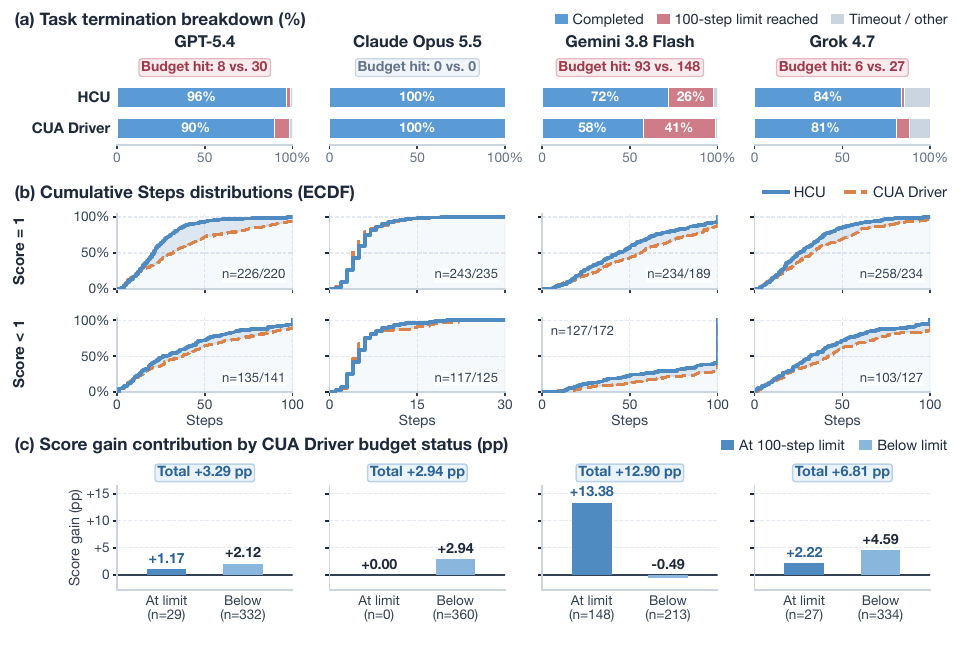}
\caption{Termination, step distributions, and limit-grouped Score gains across models (Appendix~\ref{app:crossmodel}). (a) Termination shares; counts denote termination at the 100-step limit. (b) Within-group cumulative Steps distributions; $n$: HCU / CUA Driver tasks. (c) Additive Score-gain contributions (pp) grouped by whether CUA Driver reaches the 100-step limit, with mean task counts beneath each bar.}
\label{fig:termination-length}
\label{fig:budget-groups}
\end{figure}

\paragraph{Where the step limit binds.}
Under the 100-step limit, CUA Driver reaches the limit on 148 tasks with Gemini 3.8 Flash (vs.\ 93 with HCU), 29 tasks per run with GPT-5.4, 27 with Grok 4.7, and 0 with Claude Opus 5.5 (Figure~\ref{fig:termination-length}a,b). Consequently, Gemini 3.8 Flash derives its +12.90-point gain primarily from tasks where CUA Driver hits the 100-step limit, whereas GPT-5.4 and Grok 4.7 gain mainly on tasks completed within the limit (Figure~\ref{fig:termination-length}c). On an independently defined set of 82 long tasks (those where GPT-5.4 or Grok 4.7 reaches the limit under CUA Driver), Gemini 3.8 Flash gains 30.5 points per task versus 7.7 on the remaining tasks.

\begin{figure}[!htbp]
\centering
\includegraphics[width=\linewidth]{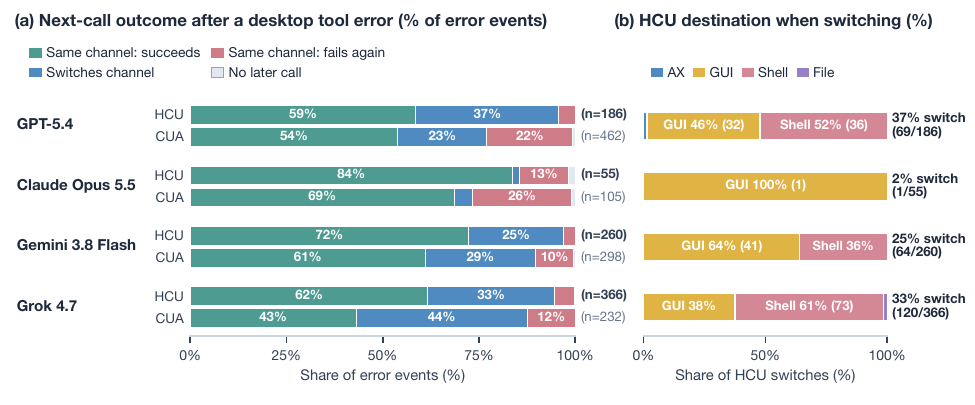}
\caption{Post-error behavior across models. (a) Next-call outcome shares over desktop-tool error events, with total error counts $n$ on the right. (b) Destination channel shares conditional on HCU switching channels.}
\label{fig:post-error}
\end{figure}

\paragraph{Post-error channel switching.}
After an AX or GUI call error, the next call switches channels in 69/186 events (37\%) for GPT-5.4, 64/260 (25\%) for Gemini 3.8 Flash, and 120/366 (33\%) for Grok 4.7, compared with 1/55 (2\%) for Claude Opus 5.5 (Figure~\ref{fig:post-error}). Claude Opus 5.5 similarly stays within CUA Driver after 100/105 errors, reflecting a model-specific preference to retry within the active channel.

\section{HCU Design and Implementation}
\label{app:implementation}
\label{app:design}

This appendix details HCU's runtime boundary, AX observation and projection pipeline, batched execution and post-action feedback, cross-channel state guards, and native platform backends.

\subsection{Interface and System Boundary}
\label{app:interface-design}
\label{app:design-implementation}

Table~\ref{tab:hcu-contract} summarizes the public operations of the \texttt{AX} and \texttt{GUI} tools. Programmatic tools (shell, code, and file operations) reside in the agent harness, while the HCU service supplies the semantic and visual desktop operations and their shared state.

\begin{table}[!htbp]
\caption{Desktop operations and main request fields.}
\label{tab:hcu-contract}
\centering\small
\begin{tabularx}{\linewidth}{@{}p{0.20\linewidth}p{0.32\linewidth}X@{}}
\toprule
Operation & Main request fields & Returned evidence and use\\
\midrule
AX \texttt{apps} & No target & List top-level windows as \texttt{wN} with title, application, and focus.\\
AX \texttt{observe} & \texttt{app} or \texttt{window\_ref}; \texttt{scope}, \texttt{root} & Return a compact window tree with state and \texttt{rN} references for subsequent calls.\\
AX \texttt{find} & \texttt{app} or \texttt{window\_ref}, \texttt{query}, \texttt{scope}; or \texttt{cursor} & Retrieve matching targets and local context without reading an entire window tree.\\
AX \texttt{act} & Up to 64 ordered \texttt{actions} and \texttt{wait} steps; \texttt{observe} & Execute steps in order, stop at the first failed step, and report per-target effects, window changes, and post-action observation.\\
GUI observation & \texttt{actions} with a single \texttt{screenshot} or \texttt{zoom} & Return the full-desktop frame, binding its window layout for subsequent coordinate checks, or an enlarged region of that frame.\\
GUI input & Up to 64 ordered \texttt{actions}: pointer, drag, scroll, text, keys, or wait & Execute mouse and keyboard input against the frame-bound layout and return the final screenshot.\\
\bottomrule
\end{tabularx}
\end{table}

\begin{figure}[!htbp]
\centering
\includegraphics[width=\linewidth]{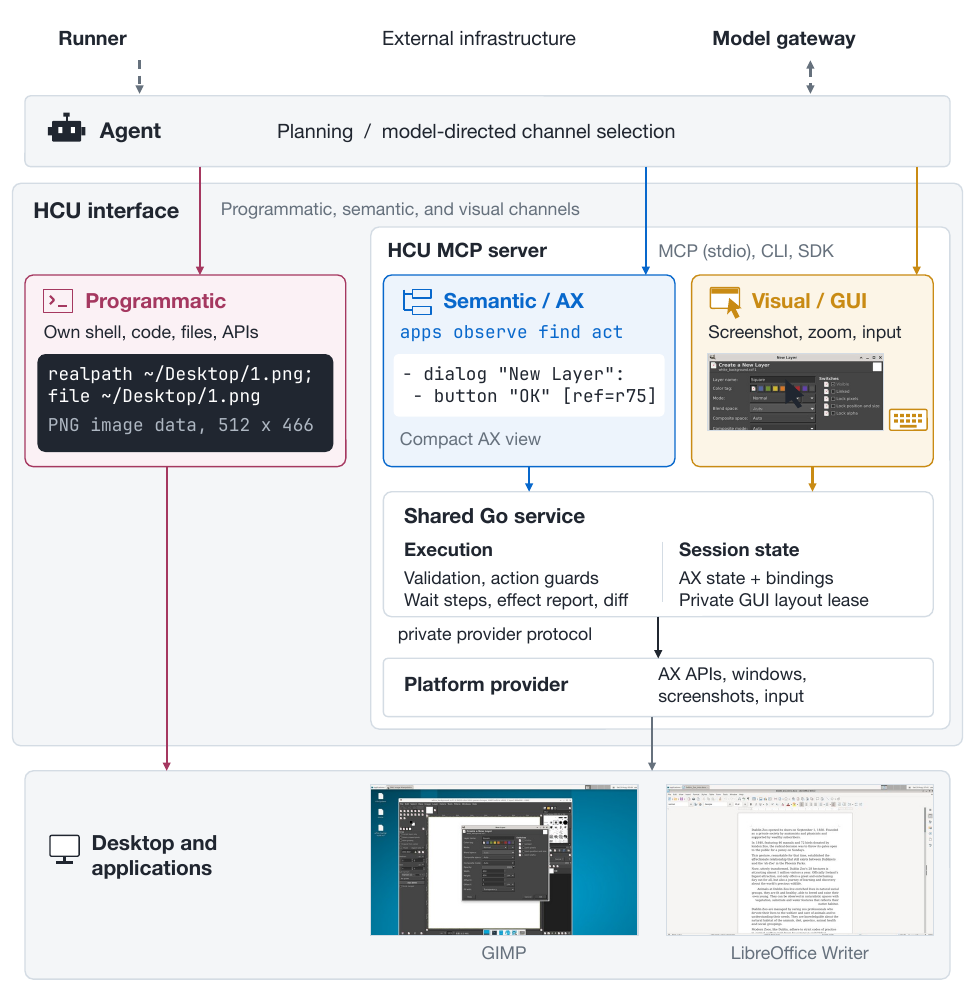}
\caption{HCU interface and MCP server boundaries. HCU unifies programmatic, semantic, and visual channels; the HCU MCP server provides the AX and GUI tools over a shared Go service, while programmatic tools live in the agent and the evaluation runner remains external. Platform backends: Figure~\ref{fig:hcu-platforms}.}
\label{fig:hcu-internals}
\end{figure}

\paragraph{Session ownership and runtime boundary.}
One process owns one desktop session (Figure~\ref{fig:hcu-internals}): either the stdio server \texttt{computer-use mcp}, or a background daemon shared by the CLI, the Python SDK, and \texttt{computer-use mcp --attach}, so target references and the input lock persist across entry points. The shared Go service validates calls and maintains AX snapshots, reference registries, find cursors, eligible diff snapshots, and the GUI layout lease. Platform providers implement tree acquisition and native input behind a typed private protocol, while the evaluation runner remains external to HCU.

\subsection{AX Observation and Target Identity}
\label{app:ax-observation}
\label{app:observations}

\paragraph{References, scope, and projection.}
The service maintains short references for applications (\texttt{aN}), logical top-level windows (\texttt{wN}), and elements (\texttt{rN}). Element actions resolve these references against live state rather than tree line numbers or display labels; \texttt{activate} and \texttt{close} also accept a window-title substring resolved to a unique \texttt{wN} before input. Window scope defaults to the full window including chrome, whereas \texttt{content} scope selects the primary content surface and \texttt{root} narrows an observation to a subtree.

From the provider's acquired records (roles, names, values, states, actions, geometry, and relations), HCU builds a typed internal tree, binds target references, and projects the compact AX view shown to the model. Projection removes empty structural wrappers, merges adjacent unreferenced text leaves, deduplicates relation-backed labels, preserves context needed to disambiguate same-name controls, displays table cell values alongside names, and summarizes repetitive sibling runs under a structural budget. Because references are bound before projection, omitting a subtree from the displayed view does not invalidate its acquired records, which remain searchable via \texttt{find}.

\begin{figure}[!htbp]
\centering
\includegraphics[width=\linewidth]{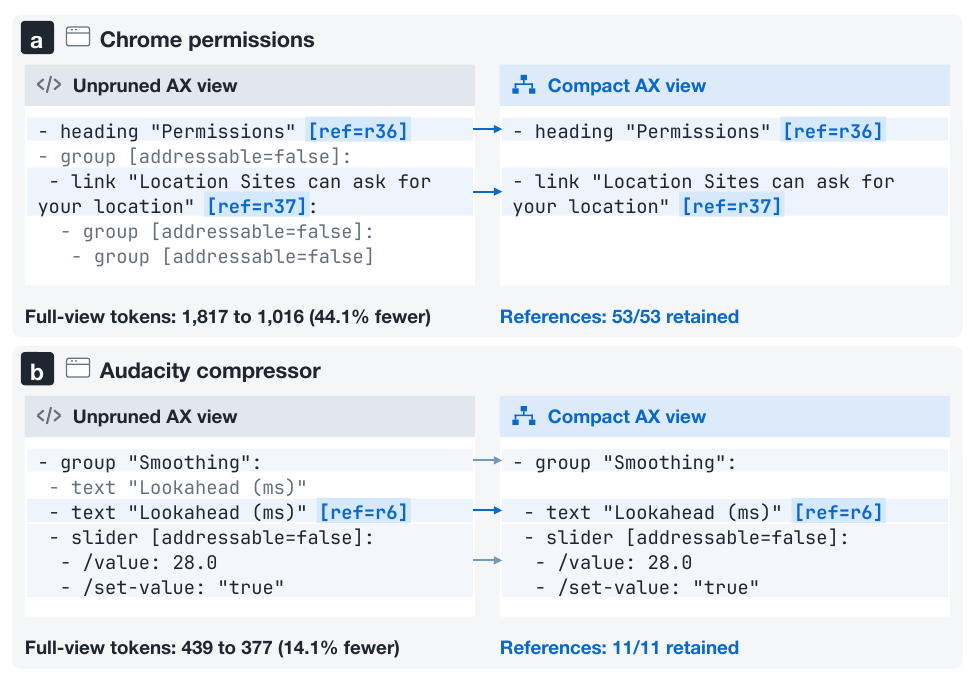}
\caption{Same-capture AX excerpts (`o200k\_base`). Counts cover complete view bodies, excluding headers and images; projection retains all issued references.}
\label{fig:projectionexamples}
\end{figure}

\begin{figure}[!htbp]
\centering
\includegraphics[width=\linewidth]{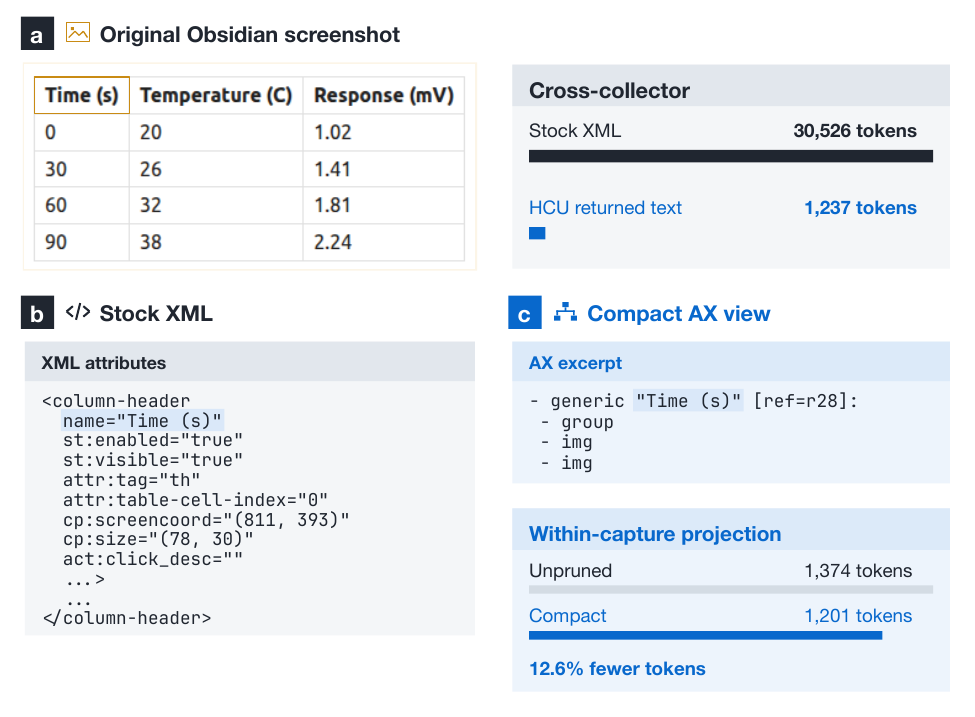}
\caption{Obsidian screenshot, stock OSWorld 2.0 XML, and compact AX excerpts.}
\label{fig:nativeobsidian}
\end{figure}

\paragraph{Same-capture AX views and stock XML comparison.}
To measure the token reduction from projection alone, we compare the delivered compact AX view against an unpruned rendering of the same acquired records across Chrome (container pruning), Audacity (label deduplication), and Obsidian (Figure~\ref{fig:projectionexamples}). Projection reduces view tokens by 44.1\% in Chrome (1,817 to 1,016 tokens), 14.1\% in Audacity (439 to 377), and 12.6\% in Obsidian (1,374 to 1,201) while retaining 100\% of issued target references (53/53, 11/11, and 48/48). Compared against the stock OSWorld 2.0 \texttt{/accessibility} XML on the same Obsidian window, the full representation drops from 30,526 tokens to 1,237 tokens (Figure~\ref{fig:nativeobsidian}).

\subsection{Action Execution and Post-Action Feedback}
\label{app:execution-feedback}
\label{app:feedback-samples}

\paragraph{Batched steps, wait steps, and feedback.}
An AX \texttt{act} call executes up to 64 ordered actions and \texttt{wait} steps within one logical window. A \texttt{wait} step sends no input and holds until a window title appears, a query matches, a referenced element reaches an expected value, state, or text, a target disappears (\texttt{gone}), or the UI changes and stops changing (\texttt{settle}). A \texttt{select} step selects an element, substring, or range between two references (\texttt{ref} to \texttt{to\_ref}) and reports the resulting selection. Action errors, wait timeouts, or window changes stop the batch immediately and report how many steps completed.

\begin{figure}[!htbp]
\centering
\includegraphics[width=\linewidth]{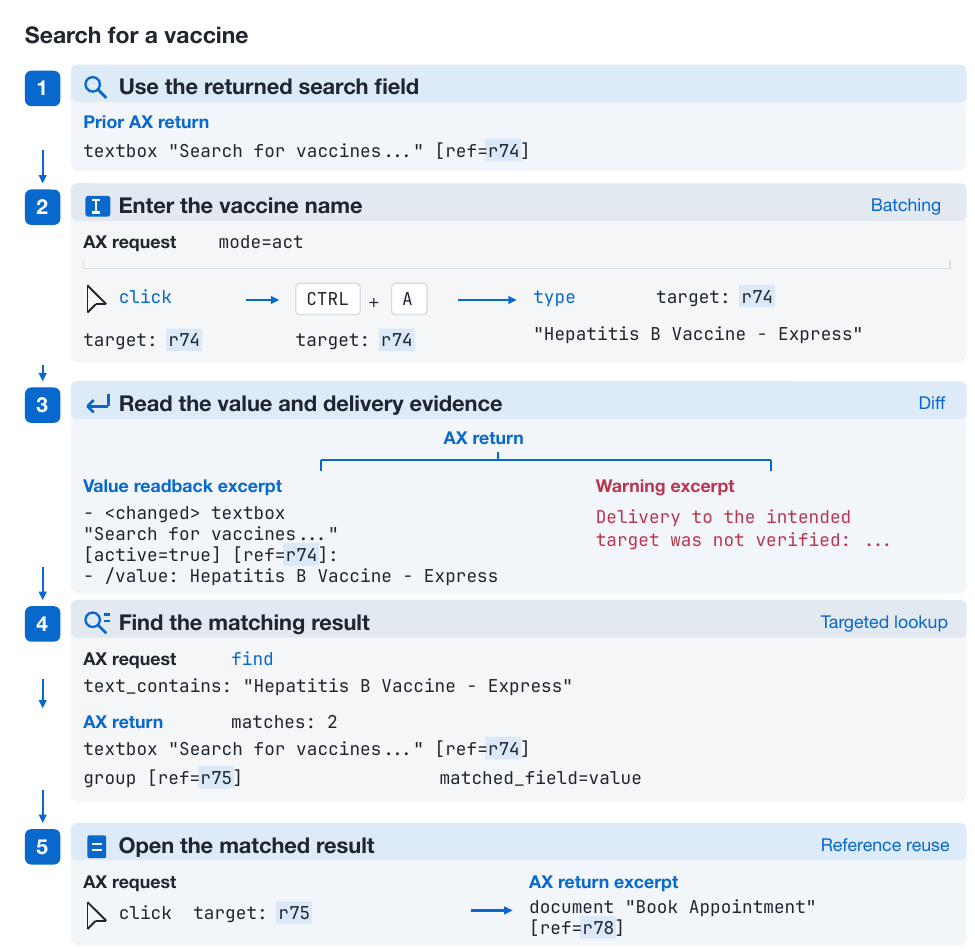}
\caption{Recorded AX interaction with GPT-5.6 Sol on OSWorld 2.0. The search batch returns the entered value, and a subsequent \texttt{find} supplies the next target.}
\label{fig:interaction}
\end{figure}

By default (\texttt{observe=diff}), an \texttt{act} call reports a per-target \texttt{effect:} line and a \texttt{windows:} line (opened, closed, or focused \texttt{wN}), followed by the structured AX diff against the last eligible snapshot delivered to the model. Every AX and GUI response starts with a JSON evidence line carrying a public error code on failure (such as \texttt{invalid\_argument}, \texttt{stale\_ref}, \texttt{gui\_frame\_stale}, \texttt{wait\_timeout}, \texttt{window\_changed}, or \texttt{effect\_unknown}), a corrected call hint, and whether input was emitted. Figure~\ref{fig:interaction} illustrates a recorded AX batch followed by \texttt{find} and continuation.

\subsection{Cross-Channel State and Failure Handling}
\label{app:state-failures}

\begin{figure}[!htbp]
\centering
\includegraphics[width=\linewidth]{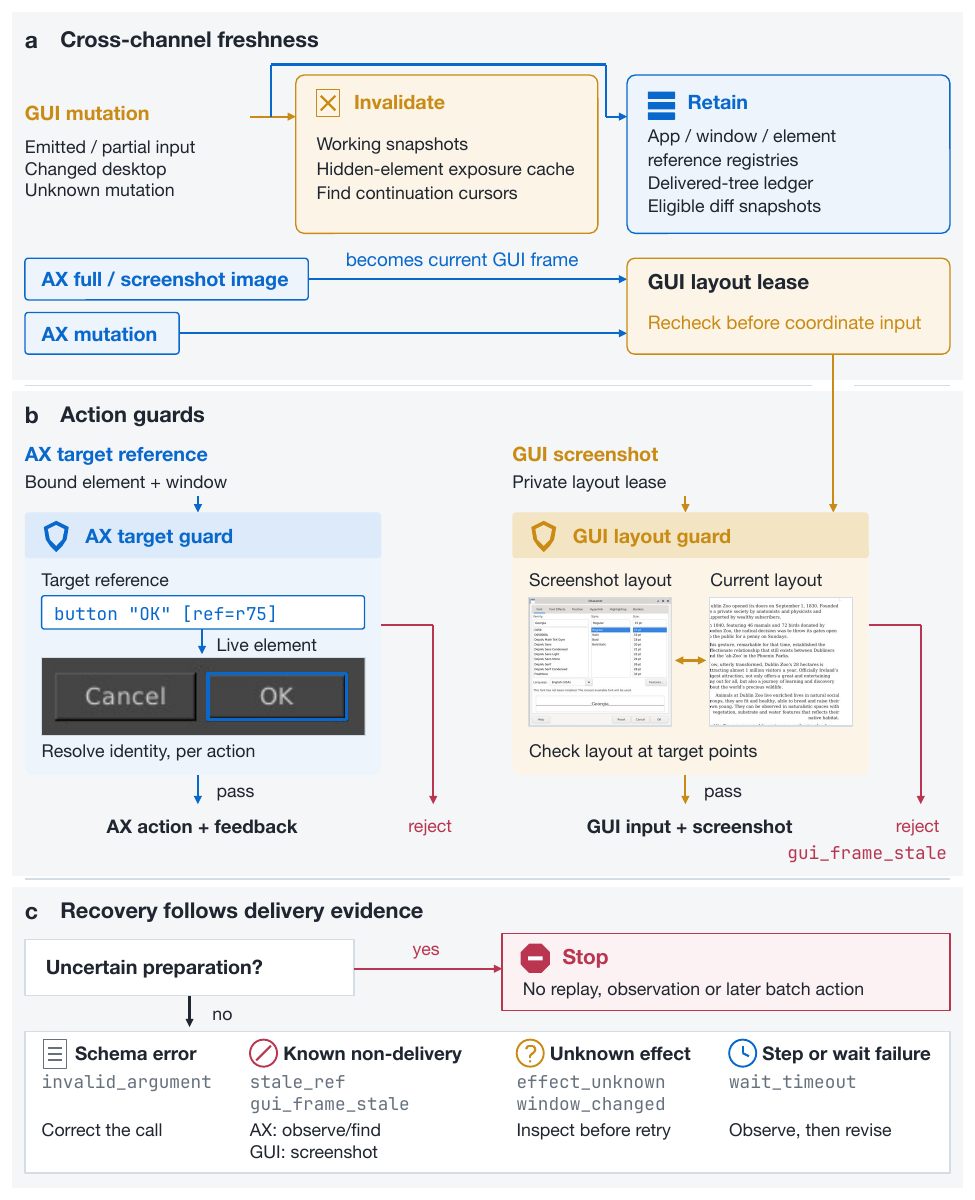}
\caption{Cross-channel state and action guards. GUI mutation invalidates working observations while retaining identities and eligible diff snapshots. The GUI guard checks window layout before coordinate input, and uncertain preparation stops the batch.}
\label{fig:cross-channel-state}
\end{figure}

\begin{figure}[!htbp]
\centering
\includegraphics[width=\linewidth]{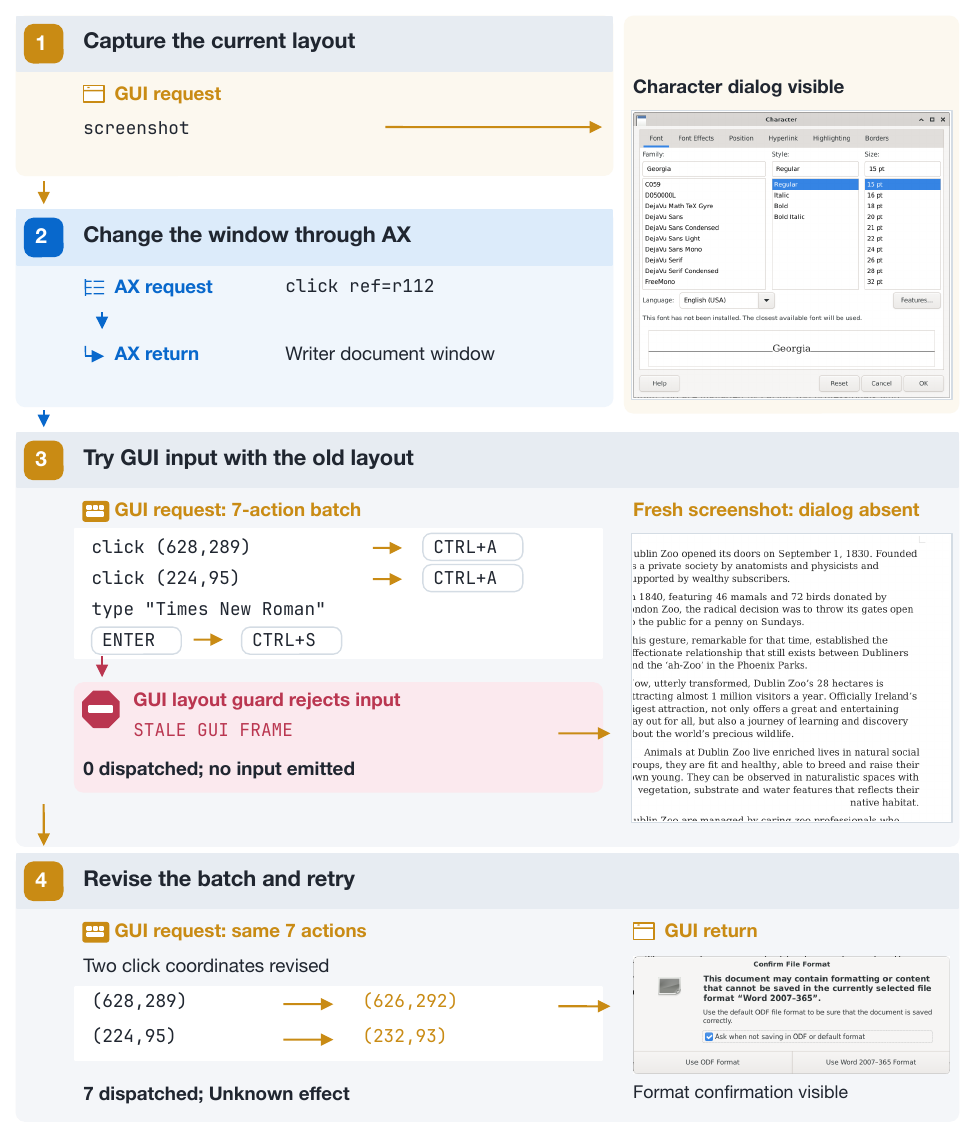}
\caption{Recorded Writer layout-guard interaction, HCU (GPT-5.4), OSWorld 1.0. The GUI batch is rejected before input and returns a fresh screenshot; the first two crops share region and scale. Retrying with revised pixel coordinates dispatches seven actions.}
\label{fig:guard}
\end{figure}

\paragraph{Cross-channel invalidation and action guards.}
Figure~\ref{fig:cross-channel-state} summarizes how HCU maintains consistency across channels. GUI input that may mutate desktop state invalidates the AX working snapshot, hidden-element cache, and find cursors while retaining reference registries and the last delivered diff snapshot. Conversely, a GUI screenshot or an AX \texttt{full}/\texttt{screenshot} observation registers the current GUI frame and its GUI layout lease. Before each AX action, the target guard resolves the reference against live state and rejects missing, ambiguous, or replaced elements (\texttt{stale\_ref}). Before each GUI coordinate action, the layout guard checks that the live window layout at the target points still matches the layout lease; on mismatch, it blocks input with \texttt{gui\_frame\_stale} and returns a fresh screenshot (recorded example in Figure~\ref{fig:guard}). If focus or pointer preparation fails in an uncertain state where a queued OS request may still fire, HCU stops desktop automation rather than replaying the batch.

\subsection{Platform Backends and Agent Integration}
\label{app:design-configuration}

\begin{figure}[!htbp]
\centering
\includegraphics[width=\linewidth]{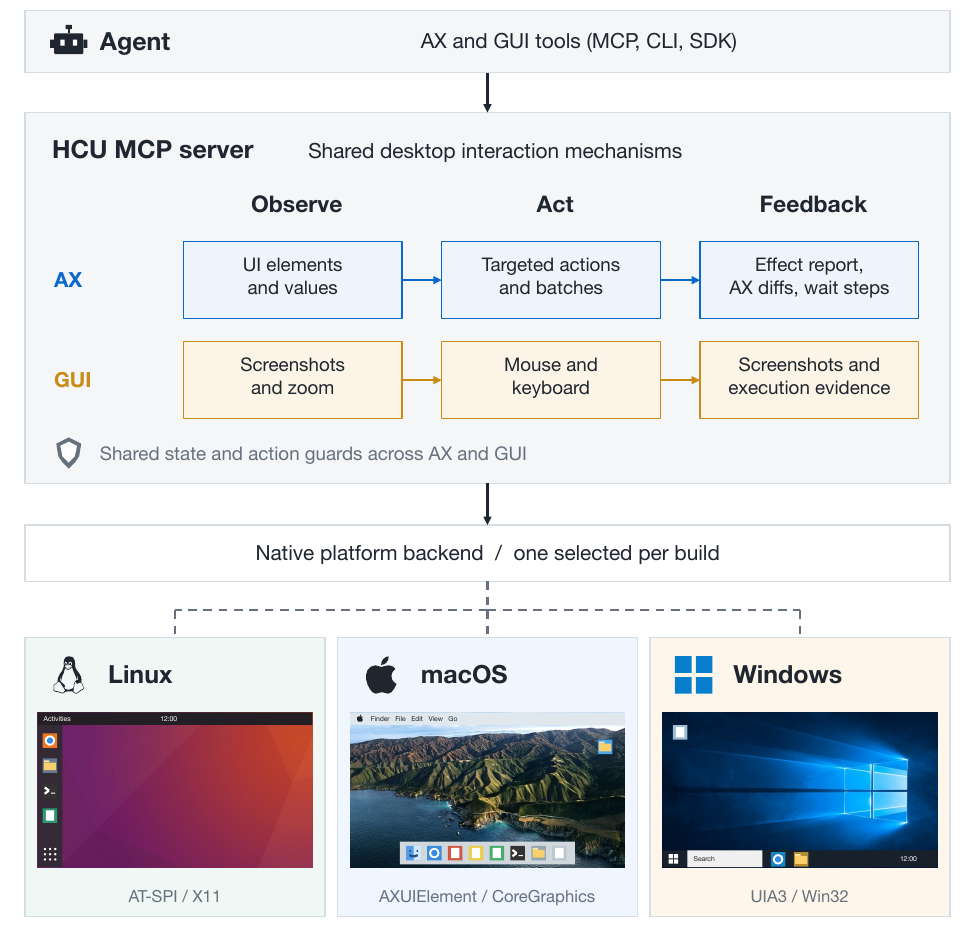}
\caption{Shared HCU MCP server and native backends. Rows show capabilities, not a fixed workflow; AX can also return screenshots, and its wait steps are optional. Dashed branches are alternative backends, one selected per build, while the agent's programmatic tools remain in the harness.}
\label{fig:hcu-platforms}
\end{figure}

HCU provides native backends for Linux, macOS, and Windows behind the same public AX and GUI contract (Figure~\ref{fig:hcu-platforms}). The shared Go service owns AX representation, reference binding, batch orchestration, wait steps, and diffs, while platform providers execute native checks, tree collection, and input: Python with AT-SPI and X11 on Linux (automatically enabling the session accessibility bus), Swift with AXUIElement, CoreGraphics, ScreenCaptureKit, and SkyLight on macOS, and C\# with UIA3 and Win32 on Windows. The binary selects its backend at build time. Standalone configuration generators and evaluation integrations register HCU alongside the built-in tools of Codex, Claude Code, Gemini CLI, and Grok Build, injecting the base policy and selected skill through each harness's native context mechanism (Appendix~\ref{app:versions}). Appendix~\ref{app:release-eval} evaluates HCU across Linux, macOS, and Windows benchmarks.

\section{Tool Definitions and Skill}
\label{app:tools}
\label{app:messages}

This appendix reproduces HCU's server instructions, \texttt{AX} and \texttt{GUI} tool descriptions, base policy (\path{instructions/base.md}), and hybrid skill (\path{instructions/hybrid.md}), followed by the two GUI action vocabularies compared in Appendix~\ref{app:gui-replacement}.

\subsection{Server Instructions and Tool Definitions}
\label{app:tool-definitions}

\begin{sourcebox}{Server instructions}
\begin{Verbatim}
Two tools for the running desktop. AX reads and operates application windows through accessibility: it lists windows, returns controls as rN refs, acts on them, and waits with wait steps. GUI takes full-desktop screenshots and sends pixel input. Use AX for named controls, values, menus, dialogs, windows, and keys. Use GUI for pixels, canvases, dragging, and controls that AX does not expose; after one targeted AX find misses a visible control, use GUI for it. Send all desktop input (window activation, keys, text, mouse) through these tools, not shell input utilities or scripted accessibility calls: the tools check the target window before input and report what changed, and outside input makes their refs and diffs stale. After GUI input, read AX again before reusing refs.
\end{Verbatim}
\end{sourcebox}

\begin{sourcebox}{\texttt{AX} tool description}
\begin{Verbatim}
Accessibility for application windows. Modes: apps lists confirmed windows (wN, title, app, active) and reports incomplete inventories; missing windows then remain unknown. observe returns the window tree with rN refs. find returns matching rN. act runs ordered steps, checks them before input, stops at the first failed step, and reports each target's change, windows opened/closed/focused, and a diff. Copy rN/wN exactly; stale refs fail. keypress/type without ref go to the focused window. select marks ref, its text substring, or ref..to_ref (paragraphs, cells) and reports the selection read back. activate/close take a title substring or wN. A wait step sends no input and fails the batch on timeout: window (title, any app), query (as find), ref with value/state/text, or text; gone inverts. settle needs no prediction: it waits until the UI changes after the step before it and stops changing; it fails only if nothing changed. After a window wait, keys go to that window, after settle to the focused one. Skip waits the report answers. Errors return code and hint. Examples:
{"mode":"act","actions":[{"type":"click","ref":"r12"},{"type":"wait","window":"Export"},{"type":"type","text":"q3.csv"},{"type":"keypress","keys":["ENTER"]}]}
{"mode":"find","window_ref":"w2","query":{"role":"button","name_contains":"Save"}}
\end{Verbatim}
\end{sourcebox}

\begin{sourcebox}{\texttt{GUI} tool description}
\begin{Verbatim}
Full-desktop screenshot and pixel input. {"actions":[{"type":"screenshot"}]} alone observes; {"type":"zoom","region":[x0,y0,x1,y1]} alone enlarges a frame region to read small text and icons. Any other batch is checked before input and returns one final screenshot. x/y are pixels of the latest full-desktop image (screen width x height) from GUI or AX full/screenshot, which return the same frame. type/keypress go to the current focus, so click the field first. Batch the actions planned from one screenshot; end the batch after an action that may open, close, or move a window, and plan the rest from the returned image. GUI cannot address windows; AX activates and closes them by title. wait is one second with no observation, at most 20 s per batch. Errors return code and hint like AX; gui_frame_stale means nothing was sent: replan from the returned screenshot. Example: {"actions":[{"type":"click","x":640,"y":412},{"type":"type","text":"hello"},{"type":"keypress","keys":["ENTER"]}]}
\end{Verbatim}
\end{sourcebox}

\subsection{Base Policy and Skill}
\label{app:skill-text}

The instruction package separates common computer-use policy in \path{instructions/base.md} from interaction guidance in \path{instructions/hybrid.md}. Companion files (\path{ax-only.md}, \path{gui-only.md}, and \path{hybrid-code.md}) adapt the same rules to single-channel ablations and code-mode harnesses.

\begin{sourcebox}{Base policy (\texttt{instructions/base.md})}
\begin{Verbatim}
# Computer-Use Principles

Complete the user's instruction end to end. Inspect real state, act within the
requested scope, and verify the delivered result. For actionable tasks, carry
the work through execution rather than stopping at a proposal. Keep progress
and final replies concise.

## Scope And Safety

- Any general-purpose route may achieve an outcome-only request. Honor an
  explicitly exclusive method, a method that is itself the user's purpose, and
  requirements about visible state. Naming an application does not by itself
  exclude an equivalent route; "leave the page open" requires the open page.
- Preserve unrelated data. Before recursive deletion, moves, overwrites, or
  permission changes, resolve the absolute targets and preview them.
- Never fabricate output, files, application state, screenshots, or
  verification. Do not read benchmark task definitions, dataset configuration,
  evaluator or gold internals, hidden checks, other tasks' traces, or
  task-derived memory, even when reachable.
- Before publishing, sending, or uploading, inspect the outgoing set for
  secrets without printing them. Remove exposed secrets only when authorized;
  if safe publication conflicts with preserving the source, ask.
- Ask only for information that the user alone can supply and that changes
  scope, safety, or an irreversible external result. Inspect evidence first,
  bundle questions, and use the configured reply channel. When the user
  invites questions and a category, range, or eligibility rule stays
  ambiguous, ask before the affected external action. For classification or
  aggregation, account for every source item and recompute dependent results
  after clarification.

## Who Owns The State

- An application that has a document open owns it. It shows its own copy and
  writes that copy on save, which overwrites edits made to the file elsewhere.
  Prefer changing an open document in that application and saving it there.
  If the application cannot reach a needed property, save and close the
  document (or confirm it has no unsaved changes), edit the file with a
  format-aware library, parse it, and reopen it.
- Change the object the user refers to. Edit it in place and keep its format;
  create a new file only when asked.
- Persist the result the way the user would: save edited documents, confirm
  settings dialogs, submit forms. Do not overwrite an original the user asked
  to keep. "Default" and "preference" mean application or user settings
  unless the user limits them to one object.

## Batch Known Actions

- Batch actions whose targets and inputs are already known and whose later
  steps do not need an intermediate result. Split at the next decision that
  needs fresh evidence; the maximum batch size is a limit, not a goal.
- Combine related shell checks or known command sequences in one call. Check
  exit status before dependent commands and stop them on failure. Run commands
  concurrently only when they are independent and do not change shared state.

## Evidence By Result Type

| Result | Evidence |
| --- | --- |
| A file | Parse the saved file |
| An exact value in an application | Read the value back from the application |
| Appearance, layout, or an image | A full-size screenshot |
| A setting or default | The effective setting after a reread, or a newly created object |
| A web change | Reload and see the page or record |
| A behavior | Exercise it once |
| Something the user should see | The window as the user will see it |

- Accepted input is not evidence. After the last change, get evidence for each
  requirement. A check returned by the changing call counts; add another
  channel only for a requirement it does not cover or when sources conflict.
- Stop when every requirement has evidence of its kind in the table. Do not
  demonstrate the result beyond that evidence: no extra test inputs,
  restarts, or feature tours. Clean up only temporary state you created that
  affects the result. Report limitations. One failed channel is
  not a blocker while another channel can do the work.
\end{Verbatim}
\end{sourcebox}

\begin{sourcebox}{Hybrid skill (\texttt{instructions/hybrid.md})}
\begin{Verbatim}
# Hybrid Computer Use

Channels: the shell (commands, scripts, format-aware libraries), the running
browser's automation endpoint when it exposes one, and the `AX` and `GUI`
tools. The tool descriptions give call shapes. This guide says which channel
fits which state, and how to combine them.

## Route By Where The State Lives

| State to change | Change it through | Reason |
| --- | --- | --- |
| Files that no running application holds open: data, code, configuration, structured documents, archives, media | Shell with a format-aware library or CLI; check the real format first | Exact, scriptable, and verifiable by parsing |
| A document an application holds open | That application: AX for menus, dialogs, and fields; GUI for what AX lacks. For a property neither reaches, save and close it, then edit the file as above | Its copy is what gets saved |
| Application or system settings | The settings UI through AX; the configuration file or CLI only when the application is closed or rereads it | A running application can write its cached settings back |
| Web pages, forms, signed-in sites | The open browser session: its endpoint if exposed, otherwise AX or GUI | Login, cookies, and tabs exist only in that session |
| Canvases, images, drawings, layout | The editor: AX for menus and dialogs, GUI for pixels and dragging | Visual objects have no names to address |
| Messages, mail, calendar items, accounts | The client application through AX | The client owns sending and sync |

- A documented automation endpoint that a running application already exposes,
  such as a browser debugging port, acts inside that application. Probe for it
  once, early. After one failed probe, use AX or GUI. Do not rebuild a site
  with shell HTTP requests, and do not start another browser or profile:
  neither shares the user's session, so their changes do not appear there.
- When a setting, property, or command name is known, change it directly.
  Otherwise learn it from the application: its menus and dialogs, or one
  narrow query of its help or manual. Do not dump whole help texts or scan
  binaries and system directories. Search the web only for facts the machine
  cannot supply; web pages often describe another version.

## Work Efficiently

- `apps` may report an incomplete inventory when an application does not
  answer. Use confirmed refs from that listing. Missing windows remain unknown;
  do not infer that they closed or that a title match is unique.
- Once the source, target, and operation are known, act. Further checks must
  answer a concrete blocker.
- Plan from one read. After `observe` or `find`, put every action you can
  already determine for that window into one `act`, while later refs and
  focus remain valid. Split when the next action needs a new observation.
- Every `act` result starts with what happened, then a diff:

```text
effect: r58 checkbox "Interface" checked=false→true
windows: opened w5 "Insert Sheet" (dialog, focused)
```

  `effect` gives each target's changed fields, `no visible change`, or
  `unknown` with the reason. `(only state flags changed in this window)` means
  no page, content, or window changed; if the control should have switched
  something, the application did not respond. `windows` names windows that opened, closed, or
  took focus, with a `wN` you can use at once. When these lines already show
  the result, do not read or wait again. Continue from the diff for other
  changes; read again only when the next decision needs more.
- One `act` can mix actions and `wait` steps; the first failed step stops
  the batch. Add a `wait` only when a later step needs a state that may
  arrive after the action returns: a slow dialog, a menu item, a value
  computed later. A checkbox that shows `checked=false→true`, or a dialog
  listed as `opened`, needs no wait.
- Wait for a known state with `window`, `query`, `ref`, or `text`. When you
  know the UI will change but not what it becomes, such as the exact title of
  the next dialog, use `{"type":"wait","settle":true}`. It waits until the UI
  changes and then stops changing, names what changed, and fails with "no
  change" only when nothing changed. Keys after it go to the window that took
  focus. Give a slow operation a longer `timeout_ms`.

```json
{"mode":"act","actions":[{"type":"click","ref":"r12"},{"type":"wait","window":"Export"},{"type":"type","text":"q3.csv"},{"type":"keypress","keys":["ENTER"]},{"type":"wait","window":"Export","gone":true}]}
{"mode":"act","actions":[{"type":"click","ref":"r7"},{"type":"wait","query":{"role":"menu item","name_contains":"Paragraph"}},{"type":"click","ref":"r9"}]}
```

- Select text or cells with a `select` step before a command that applies to
  the selection: `{"type":"select","ref":"r4","to_ref":"r5"}` selects from
  paragraph `r4` through paragraph `r5`, or the cell range between two cells;
  `"text"` selects one substring of `ref`. The result reports what is now
  selected; check it before applying the command.
- Take a GUI screenshot when the question is visual or AX lacks a visible
  element. Do not alternate screenshots and AX reads of an unchanged state.
  An AX `observe:"full"` image is the same full-desktop GUI frame; plan GUI
  coordinates from it without another screenshot. A GUI `zoom`
  enlarges a frame region to read small text or icons; click with the frame
  pixels its mapping gives.
- Put GUI actions determined from the same screenshot into one `actions`
  array when focus and target positions remain valid. A call returns only a
  final screenshot. End the batch when a window or dialog opens, closes,
  moves, resizes, switches, or navigates; use the returned screenshot to plan
  subsequent coordinates. A screenshot action must stand alone.
- On Linux, start a GUI application from the shell with
  `setsid -f APP ARGS >/tmp/APP.log 2>&1 </dev/null`. Some agent shells kill
  the process group of each command when it returns; an application started
  there with `&`, `nohup`, or `disown` dies without showing a window.
- On macOS, use `open -a APP` to launch through LaunchServices. Reuse an
  existing application instance and verify its window before sending input.

## When Something Fails

Every AX and GUI error starts with a `code`, a recovery `hint`,
and whether input was sent. Read `failure_code`, `input_emitted`,
`completed`, and `retry_same_request` when present. If input was sent or
`completed` is nonzero, inspect state and send only what remains.
Never repeat the request when `retry_same_request=false`. Next step by code:

- `invalid_argument`: nothing was sent. Fix the field named in `path` as the
  hint shows. The control is not missing.
- `stale_ref`: drop every ref from that result. Use refs from a newer result,
  or one fresh `find`.
- `window_changed`: the batch stopped after `completed` actions. Read the
  returned state and send only what remains.
- `wait_timeout`: the batch stopped at that `wait`; the steps before it ran
  and none after it. The message says what was seen instead, or that the
  state could not be read. Read the returned state and revise the plan; do
  not resend the same batch.
- `effect_unknown`: input may have been sent. Inspect state before any retry,
  so that nothing is applied twice.
- `ambiguous_window`, `window_not_found`: choose a `wN` from the listed windows.
- `no_focused_window`: start the batch with `activate`.
- `app_unavailable`: if the application is still starting, wait once for
  its window with `{"mode":"act","actions":[{"type":"wait","window":"..."}]}`;
  otherwise use GUI for it.
- `permission_denied`: restore the provider's operating-system permissions
  and restart it before observing again. Repeating a screenshot cannot grant
  access. If earlier input was sent, inspect state before sending more.
- `runtime_error`: use the specific failure reason. For a temporary read
  failure, retry the read once unless the result forbids retrying. Inspect
  current state before retrying input. For `coordinate_target_unverified`,
  take a GUI screenshot and inspect covering windows and the target;
  refreshing refs alone does not fix screen-point ownership.
- A zero-result find or a timed-out wait may list nearby elements or window
  refs. Check those suggestions against the intended target before retrying.
- `gui_frame_stale`: the window under a GUI coordinate changed after its
  screenshot, so nothing was sent. Plan again from the returned screenshot.

A stop response ends desktop automation: report the uncertainty
and stop.
After two attempts on the same hypothesis without new evidence, change the
mechanism or the channel. Do not nudge coordinates, add Enter, add waits, or
relaunch applications to escape uncertainty.
\end{Verbatim}
\end{sourcebox}

\subsection{GUI Action Vocabularies}
\label{app:gui-action-space-defs}

Table~\ref{tab:gui-action-spaces} lists HCU's default GPT-style GUI vocabulary and the Claude-style vocabulary evaluated in the action-space replacement of Appendix~\ref{app:gui-replacement}.

\begin{table}[!htbp]
\caption{GUI action vocabularies. The GPT-style space is HCU's default.}
\label{tab:gui-action-spaces}
\centering\small
\begin{tabularx}{\linewidth}{@{}l>{\raggedright\arraybackslash}X>{\raggedright\arraybackslash}X@{}}
\toprule
Operation & GPT-style (\texttt{type} field) & Claude-style (\texttt{action} field)\\
\midrule
Observe & \texttt{screenshot}; \texttt{zoom} of a frame region & \texttt{screenshot}, \texttt{zoom}, \texttt{cursor\_position}\\
Click & \texttt{click} with \texttt{button} and modifier \texttt{keys}; \texttt{double\_click} & \texttt{left\_click}, \texttt{right\_click}, \texttt{middle\_click}, \texttt{double\_click}, \texttt{triple\_click}\\
Pointer & \texttt{move}; \texttt{drag} along a \texttt{path} & \texttt{mouse\_move}; \texttt{left\_click\_drag}, \texttt{left\_mouse\_down}, \texttt{left\_mouse\_up}\\
Scroll & \texttt{scroll} with \texttt{scroll\_x} and \texttt{scroll\_y} deltas & \texttt{scroll} by direction, in wheel clicks\\
Keyboard & \texttt{type}; \texttt{keypress} with a key array & \texttt{type}; \texttt{key} with a key-chord string; \texttt{hold\_key}\\
Wait & \texttt{wait}, one second & \texttt{wait}\\
\bottomrule
\end{tabularx}
\end{table}

\end{document}